\documentclass[showpacs,aps,prl,twocolumn,superscriptaddress]{revtex4-1}

\usepackage{graphicx}  
\usepackage{subfigure}
\usepackage{dcolumn}   
\usepackage{bm}        
\usepackage{amssymb}   
\usepackage{comment}
\usepackage{hyperref}
\usepackage{color}
\usepackage{amsmath}
\usepackage{mathrsfs}
\usepackage{mathcomp}
\usepackage{textcomp}
\usepackage{dsfont}
\usepackage{esint}
\usepackage{braket}
\usepackage{textgreek}
\usepackage{lipsum}

\begin{document}

\title{Tomonaga-Luttinger liquid and localization in Weyl semimetals}

\author{Xiao-Xiao Zhang}
\affiliation{Department of Applied Physics, The University of Tokyo, 7-3-1 Hongo, Bunkyo-ku, Tokyo 113-8656, Japan}
\author{Naoto Nagaosa}
\affiliation{Department of Applied Physics, The University of Tokyo, 7-3-1 Hongo, Bunkyo-ku, Tokyo 113-8656, Japan}
\affiliation{RIKEN Center for Emergent Matter Science (CEMS), 2-1 Hirosawa, Wako, Saitama 351-0198, Japan}


\newcommand\dd{\mathrm{d}}
\newcommand\ii{\mathrm{i}}
\newcommand\ee{\mathrm{e}}
\newcommand\zz{\mathtt{z}}
\makeatletter
\let\newtitle\@title
\let\newauthor\@author
\def\ExtendSymbol#1#2#3#4#5{\ext@arrow 0099{\arrowfill@#1#2#3}{#4}{#5}}
\newcommand\LongEqual[2][]{\ExtendSymbol{=}{=}{=}{#1}{#2}}
\newcommand\LongArrow[2][]{\ExtendSymbol{-}{-}{\rightarrow}{#1}{#2}}
\newcommand{\cev}[1]{\reflectbox{\ensuremath{\vec{\reflectbox{\ensuremath{#1}}}}}}
\newcommand{\blue}[1]{\textcolor{blue}{#1}} 
\newcommand{\mycomment}[1]{} 
\makeatother

\begin{abstract}
We study both noncentrosymmetric and time-reversal breaking Weyl semimetal systems under a strong magnetic field with the Coulomb interaction. 
The three-dimensional bulk system is reduced to many mutually interacting quasi-one-dimensional wires. Each strongly correlated wire can be approached within the Tomonaga-Luttinger liquid formalism. Including impurity scatterings, we inspect the localization effect and the temperature dependence of the electrical resistivity. The effect of a large number of Weyl points in real materials is also discussed.
\end{abstract}
\keywords{}

\maketitle



\textit{Introduction.---}%
The realization of linear band crossings in three dimensions (3D) in the Weyl semimetals are sparking keen interests\cite{TaAS1,*TaAS2}. 
This lends credence to the concept of Weyl fermion\cite{Weyl1929} in the context of variou condensed matter systems\cite{Weyl2007,Weyl2011}. In principle, any solid-state realization should bear time-reversal symmetry breaking (TRB) and/or inversion symmetry breaking (IB)\cite{WeylwithT1,WeylwithP1,WeylwithP2,AHE2,predictCME} so as to lift the Kramers degeneracy and to generate nonzero Berry curvatures. 
The Weyl point is interesting as a 3D counterpart of the two-dimensional (2D) Dirac physics\cite{DiracFermion1,*DiracFermion2}, 
which means topologically protected monopoles of the momentum-space Berry phase\cite{Volovik}\mycomment{\cite{EEMF0,EEMF2}}. Among others, the chiral magnetic effect\cite{CME0,*CME1,*CME2} as a result of the chiral anomaly\cite{Adler,*Bell&Jackiw,Nielson-Ninomiya1,*Nielson-Ninomiya2,ReviewQi,thetaWeyl,*ReviewBurkov} is observed as negative magnetoresistance in Dirac/Weyl semimetals\cite{seeCMEDirac2,*seeCMEDirac1,seeCMEWeyl1,*seeCMEWeyl2} once the chiral imbalance of chemical potential is generated by parallel electric and magnetic fields.

The intriguing facet of the magnetotransport in Dirac/Weyl semimetals mainly comes from the the unique Landau level formation dissimilar to that of quadratic electronic bands, where the lowest Landau level, a linearly dispersed chiral mode along the direction of the magnetic field, is well separated from the higher levels by a cyclotron gap $\propto\sqrt{B}$, whose 2D variant has been vastly explored in graphene\cite{DiracLandauLevel1,*DiracLandauLevel2,*DiracLandauLevel3,*graphene}. A further stage is when the (ultra) quantum limit is achieved\cite{massgeneDiracSM1,*massgeneDiracSM2}, enabling the lowest Landau level to play a major role in shaping the low-energy physics. In this limit, the magnetic length $l_B=1/\sqrt{eB}$ (setting $\hbar=1$) becomes shorter than the Fermi wavelength since the quantized orbit of electrons shrinks with an increasing $B$ and the lowest Landau level possesses the majority of population\cite{bismuthQL}. Remarkably, it implies a field-induced dimensional reduction\cite{DimensionalReduction0,*DimensionalReduction1} that will strongly enhance correlations hence the advent of the (quasi-) 1D system without electron quasiparticle excitations. This connects to the long-lasting search or application of the Tomonaga-Luttinger liquid (TLL) physics\cite{old1D,*transportLLreview,*1DRMP,*1D21st}, including semiconductor quantum wires\cite{semicondLL1,*semicondLL2,FQHE1Dedge0}, single-walled carbon nanotubes\cite{LLCNT1,*LLCNT2,CoulombCNT1,*CoulombCNT2}, edge states in fractional quantum Hall states\cite{FQHLL1,*FQHLL2,*FQHLL3,FQHE1Dedge1,*FQHE1Dedge2} and 2D topological insulators\cite{FQHE1Dedge3,2DTCIbosonize}, and so on.

Because of the large cyclotron gap, it is expected and confirmed that the Dirac/Weyl semimetals can be driven to the quantum limit at lower magnetic fields than semiconductors\cite{massgeneDiracSM1,massgeneDiracSM2}. Due to the instability from electron correlations, one possibility is the gap-opening or dynamical mass generation\cite{gap_opening1,*gap_opening2} in the nominally massless semimetal as density waves are formed\cite{Ran_model,*WeylSDW,*WeylCDW1,*WeylCDW2}. Instead, in this study we will explore a different scenario for Weyl semimetals at the magnetic quantum limit. Two minimal models are considered and shown to be closely related, corresponding to the predicted TRB pyrochlore iridates\cite{Weyl2011} and the realized nonmagnetic and IB transition metal monoarsenides/monophosphides\cite{TaAS1,*TaAS2,TaAS3,*TaAS4,NbAs1,TaP1,*TaP2,NbP1,*NbP2,*NbP3,*NbP4,
predict2,*predict1,predict3,*predict4}. We incorporate long-range Coulomb interactions and show how the TLL state naturally emerges as a result of singling out the chiral 1D channels by applying a magnetic field. Adopting the coherent state basis of Landau levels, the 3D system is transformed into a lattice of parallel quasi-1D wires interacting with each other. Focusing on the on-wire effective model, we investigate the localization effect due to impurity scatterings. To facilitate experimental investigations, we derive the temperature dependence of resistivity and show how the relatively large number of Weyl points in materials affects the properties.





\begin{figure}
  \scalebox{0.38}{\includegraphics{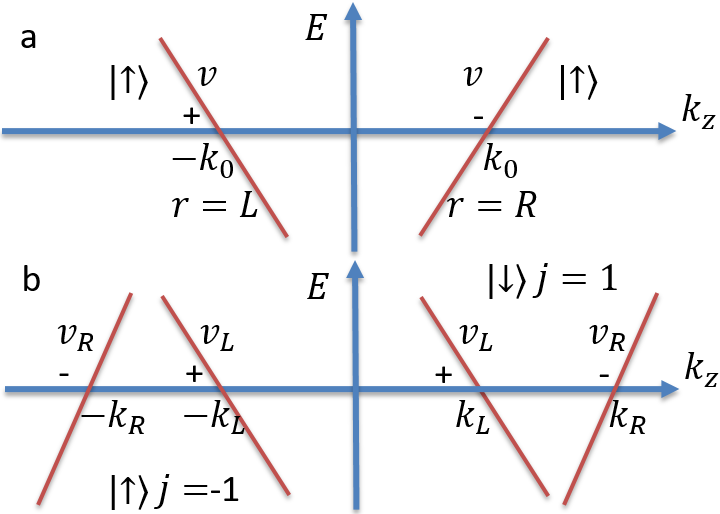}}
  \caption{1D linear dispersions along $k_z$. There are two [four] Weyl points at $\pm k_0$ [$\pm k_R,\pm k_L$] with topological charges denoted by $\pm$ for the (a) TRB [(b) IB] case. We label 1D channels with chirality index $r=R/L$, written as $r=\pm 1$ in calculation, in both cases and side index $j=\pm 1$ only in (b). $\ket{\uparrow/\downarrow}$ are pseudospin states. Two modes at each side in (b) have unequal Fermi velocities $v_R \neq v_L>0$ in general.
  }  \label{fig:1Dmodes}
\end{figure}

\textit{Weyl semimetals under strong magnetic field.---}
We start from two minimal lattice models of the form $h(\vec{k})=\sum_i{d_i}\sigma_i$ with psuedospin $\sigma_i$, realizing the one-pair TRB and the two-pair IB cases with $d_x=\sin{k_x}\sin{k_z}\,, d_y=\sin{k_y}\sin{k_z}\,, d_z=(\cos{k_z}-\cos{k_0})-2(2-\cos{k_x}-\cos{k_y})$ and $d_x=\sin{k_x}\sin{k_z}\,, d_y=\sin{k_y}\,, d_z=-(\cos{k_z}-\cos{k_L})(\cos{k_z}-\cos{k_R})-2(2-\cos{k_x}-\cos{k_y})$, respectively. 
Based on the Landau quantization solution of a Weyl Hamiltonian under magnetic field $\vec{B}=B\hat{z}$ [Supplemental Material (SM)\cite{SM}\mycomment{~\ref{App:WSMmodel}}], one can obtain the 1D linear modes in Fig.~\ref{fig:1Dmodes}, wherein any two modes 
of the same value of velocity are related by inversion or time-reversal symmetry in the TRB and IB cases, respectively. 
As is mentioned below, the TRB case can be directly mapped to part of the more complex IB case, we henceforth focus on the latter unless otherwise stated and use the shorthand channel index $\kappa=(j,r)$.

As a matter of fact, the Weyl points are not necessary to reside along a single line for a IB material realization. Nonetheless, the model is adequate to illustrate the key features. The situation in Fig.~\ref{fig:1Dmodes} is general for a Weyl semimetal up to some reversal of chiralities and pseudospins. And our theory does not directly rely on this because in the Landau level solutions, positions of Weyl points in $k_x$-$k_y$ plane do not enter the 1D dispersions but the spatial-part wavefunctions, reflecting the large degeneracy. For the 1D modes, it is the momentum parallel to the quantizing field that matters.

\textit{Coulomb interaction.---}We then need to find the scattering processes. Except from excluding pseudospin-flip scatterings, the merit of the long-range Coulomb interaction mainly consists in that the scatterings accompanied by large momentum transfer are negligibly small than those with nearly zero transfer. Therefore, we can take into account four types of forward scatterings without momentum transfer directly connecting distinct 1D modes, viz., the Coulombic scatterings $\braket{\kappa,\kappa_i\vert \hat{U} \vert \kappa,\kappa_i}\psi_\kappa^\dag \psi_{\kappa_i}^\dag \psi_{\kappa_i}\psi_\kappa$ for a generic $\kappa=(j,r)$ with $\kappa_i$ running over $(j,r)\,,(j,-r)\,,(-j,r)\,,(-j,-r)$ for $i=1,2,3,4$.
This, however, overlooks the Landau level degeneracy, which also plays an important role since the interaction depends on both the energy dispersions and the wavefunctions. In other words, each Weyl point, under a magnetic field $B$, yields not one but many more linear modes of the number of degeneracy proportional to $B$, which are identical to the ones shown in Fig.~\ref{fig:1Dmodes}.
The degenerate subspace hereof can be expanded using the over-complete set of coherent state basis\cite{LandauLevel1,*LandauLevel2,CoherentState3,*Glazman}, which is constructed from the spatially localized ground state wavefunction $\chi_{\vec{R}=0}(\vec{r}_\perp)=\frac{1}{\sqrt{2\pi}l_B}\ee^{-\vec{r}_\perp^2/4{l_B}^2}$ by displacing its center of orbit (guiding center) $\vec{R}$ along a 2D square lattice with spacing $\sqrt{2\pi}l_B$ where $\vec{r}_\perp$ is the coordinate in the $x$-$y$ plane.

As detailed in SM\cite{SM}\mycomment{~\ref{App:Coulomb}}, this approach to formulating electron-electron correlations defined in real space provides us an intuitive and transparent picture. Now electrons at these coherent states are localized around the guiding centers in the $x$-$y$ plane but relatively unconstrained to move along the magnetic field ($z$-axis), giving rise to many mutually parallel quasi-1D wires of the number of degeneracy threading the 2D lattice of $\vec{R}$. Each wire inherits four 1D modes in Fig.~\ref{fig:1Dmodes}(b). The salient point is that the scattering processes mentioned above, which remain intact although, can now have both inter-wire and intra-wire ones. One can then express the electron field $\Psi(\vec{r})=\sum_{\vec{R}\kappa}{\chi_{\vec{R}}(\vec{r}_\perp)\psi_{\kappa\vec{R}}(z)\beta_\kappa}$ where $\beta_\kappa$ is the pseudospin wavefunction. The noninteracting Hamiltonian for all the 1D modes is therefore given by $H_0 = \sum_{\kappa k_z\vec{R}}{\varepsilon_\kappa(k_z) \psi_{\kappa\vec{R}}^\dag(k_z) \psi_{\kappa\vec{R}}(k_z)}$ with $\varepsilon_\kappa(k_z)=rv_r(k_z-jk_r)$. In the limit of strong magnetic fields $l_B\rightarrow0$, using the asymptotic orthogonality of the coherent states\cite{CoherentState3,Glazman}, the interaction part takes the form $H_I = \frac{1}{2} \sum_{\vec{R}\vec{R}'\kappa\kappa'} \int\dd z\dd z' \frac{k_e e^2/\varepsilon_r}{\sqrt{(z-z')^2+(\vec{R}-\vec{R}')^2}} \times\psi_{\kappa\vec{R}}^\dag(z)\psi_{\kappa\vec{R}}(z)\psi_{\kappa'\vec{R}'}^\dag(z')\psi_{\kappa'\vec{R}'}(z')$ wherein Coulomb's constant $k_e=\frac{1}{4\pi\varepsilon_0}$, vacuum (relative) permittivity $\varepsilon_0$ ($\varepsilon_r$), and $\vec{R}$ replaces $\vec{r}_\perp$ in the potential because of the transverse confinement at strong fields.

\textit{Charge-chirality separated bosonization.---}Next we bosonize this system of many interacting quasi-1D wires\cite{Bosonization1,GiamarchiBook}.
As shown in Fig.~\ref{fig:1Dmodes}(b), the opposite-chirality modes do not share the same velocity, which is a bit unorthodox for conventional bosonized fields combining two chiralities. The physically transparent way out is to start from the chiral boson field $\varphi^\kappa$ that bosonizes a single Weyl fermion in $(1+1)$-dimensions, i.e., any on-wire linear mode in Fig.~\ref{fig:1Dmodes}(b) is expressed as $\psi_{\kappa\vec{R}}(z)=\Upsilon_\kappa\frac{1}{\sqrt{2\pi\alpha}}\ee^{\ii jk_rz}\ee^{\ii r\varphi_{\vec{R}}^\kappa(z)}$
in which $\alpha$ is the lattice cutoff and $\Upsilon$ is the Klein factor (omitted henceforth). Resembling the standard spin-charge separation, it is convenient to separate the charge and chirality degrees of freedom $\vec{\zeta} = (\theta_\rho,\theta_\chi,\phi_\rho,\phi_\chi)^\mathrm{T}
=\frac{1}{2}
\begin{bmatrix}
H & -H \\
-H & -H 
\end{bmatrix}
(\varphi^{\kappa_1},\ldots,\varphi^{\kappa_4})^\mathrm{T}
\mycomment{(\varphi^{j,r},\varphi^{j,-r},\varphi^{-j,r},\varphi^{-j,-r})^\mathrm{T}}$
where $H$ is the Hadamard matrix.
 This block-diagonalizes the action matrix in $\mathcal{S} = \frac{1}{2\pi\beta A_\perp V} \sum_p \vec{\zeta}^\dag_p M_p \vec{\zeta}_p$ such that $M_p=\frac{q^2}{4}\mathrm{diag}(M_{\theta,p},M_{\phi,p})$ with $M_{\theta,p} = \begin{bmatrix}
v^+ & v^--\frac{2\ii\omega}{q} \\
v^--\frac{2\ii\omega}{q} & v^+ 
\end{bmatrix}\,,
M_{\phi,p} =  \begin{bmatrix}
8V_g+v^+ & v^--\frac{2\ii\omega}{q} \\
v^--\frac{2\ii\omega}{q} & v^+ 
\end{bmatrix}$
wherein $v^\pm=v_R\pm v_L$, $V_g=\frac{2k_e e^2}{\varepsilon_r A_\perp k^2}$, $A_\perp=2\pi {l_B}^2$ is the area of a unit cell of the guiding center lattice and $r=(z,\vec{R},\tau)=(\vec{r},\tau)$ in real space with the corresponding $p=(q,\vec{Q},\omega)=(\vec{k},\omega)$ in energy-momentum space. We assume the total volume of the system $V=\Omega\Omega_\perp$ with the volumes of $\hat{z}$ direction and $x$-$y$ plane being $\Omega$ and $\Omega_\perp$, respectively.
Note that the Coulomb interaction enters the $\phi_\rho$-quadratic term since $\phi_\rho$ is directly related to the total particle density $\rho = -\frac{1}{\pi} \nabla\phi_\rho$. Following a similar flow of construction, one observes that the TRB case with two degrees of freedom, using the standard $(\phi,\theta)$ fields with $\varphi^r=-(\phi-r\theta)$, has an action exactly mapped from the previous $M_{\phi,p}$ with $(V_g,v^+,v^-,\omega) \rightarrow (V_g,4v,0,-2\omega)$.

\begin{figure}
\scalebox{0.39}{\includegraphics{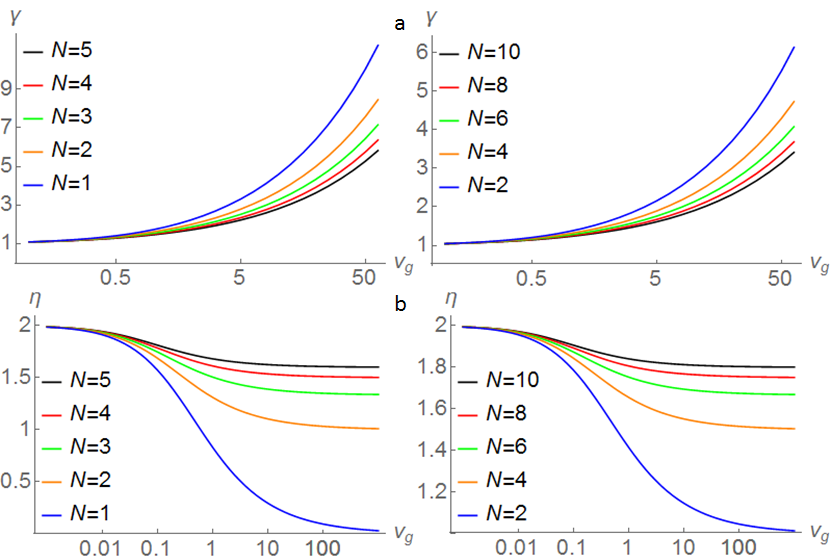}}
  \caption{(Color online) Exponents (a) $\gamma$ and (b) $\eta$ depend on $v_g$ and the number $N$ of opposite-chirality Weyl-point pairs. Left: TRB case ($v=1$). Right: IB case ($v_R=3,v_L=1$). Fermi velocities and $v_g$ are in units of $\frac{k_e e^2}{h}$.}\label{fig:multiExponent}
\end{figure}

We calculate the electron Green's function $\mathcal{G}_\kappa(z,\tau) = -\braket{\mathrm{T}_\tau \psi_\kappa(r)\psi_\kappa^\dag(0)}$ of an on-wire 1D mode $\kappa$ for $r=(z,\vec{0},\tau>0)$\cite{SM}\mycomment{~\ref{App:GreenFunc}}.
At the non-interacting limit, it reduces to the free Green's function $\mathcal{G}_\kappa(z,\tau) 
= -\frac{\ee^{\ii jk_rz}}{2\pi\alpha} \left[ \frac{\alpha+v_r\tau-\ii r z}{\alpha} \right]^{-1}$. For the long-distance 
asymptotic behavior of the equal-time correlation, $\mathcal{G}_\kappa(z,0) \sim z^{-\gamma}$ where, as shown in Fig.~\ref{fig:multiExponent}(a), $\gamma$ increases from unity with $v_g=\frac{k_e e^2}{2\pi\varepsilon_r}$ \mycomment{that replaces $V_g$ in the effective model below and characterizes} characterizing the material-dependent (via $\varepsilon_r$) strength of Coulomb interaction. Therefore, the correlation decays faster than a free one as expected for a TLL since single-particle excitations are suppressed.

\textit{Localization in an effective 1D wire.---}
From now on, in order to take a direct look at the 1D physics emerged as a result of the strong magnetic field, we derive an effective model for a particular wire. In the path-integral formalism, aided by auxiliary Lagrange multiplier fields\cite{Nagaosa2}\mycomment{\cite{NagaosaSingleImp,Nagaosa2}}, one can integrate out all the other fields except the ones on the wire of interest and thus arrives at the 1D effective action $\mathcal{S}_\mathrm{1D} = \frac{1}{2\pi\beta\Omega} \sum_{\vec{q}}\vec{\zeta}_{\vec{q}}^\dag \mathcal{M}_{\vec{q}} \vec{\zeta}_{\vec{q}}$ where $\vec{q}=(q,\omega)$, $\mathcal{M}_{\vec{q}}=\mathrm{diag}(\mathcal{M}_{\theta,\vec{q}},\mathcal{M}_{\phi,\vec{q}})$ with $\mathcal{M}_{\phi,\vec{q}}$ becoming complicated\cite{SM}\mycomment{~\ref{App:Eff1Dmodel}}. Because of integrating out $\vec{Q}$ up to the Brillouin zone boundary such that ${Q^*}^2A_\perp=4\pi$, $V_g$ in a way becomes a renormalized $v_{g'}=v_g \ln{\left[1- \frac{(\ii\omega-q v_1)(\ii\omega +q v_{-1})}{2 q^2v^+ v_g}\right]}$, appearing especially in the estimation of the exponents. Certainly, these complexities result from the Coulomb interaction between the wire of interest and all the other. 
For a purely 1D system with a long-range interaction, the dimensionless Luttinger parameter $K_\rho$, which includes the interaction effects for the charge sector, would effectively tend to zero due to the long-range divergence, leading to a slower decay than power law in the correlation functions.
However, in our system, the presence of many 1D wires resultant from the large degeneracy screens the Coulomb interaction and will not suffer from a similar divergence\cite{CoulombQuasi1D1,*CoulombQuasi1D2,*CoulombQuasi1D3}. Indeed, denser packing of the wires gives rise to a larger screening effect as seen when we discuss the multi-pair case.

In 1D, the effects of disorder and interaction are both enhanced, and the resultant localization of electrons should be much pronounced and manifest in observable quantities. Firstly, backward scatterings without reversing the side index are permissible since we do not consider the impurity potential to alter the pseudospin state, i.e., one has the scattering term $H_\mathrm{imp} = \int \dd z \tilde{\mathcal{V}}(z)\sum_j{\psi_{jR}^\dag(z)\psi_{jL}(z)+\psi_{jL}^\dag(z)\psi_{jR}(z)}$. After bosonization, this becomes $H_\mathrm{imp} = \int \dd z \mathcal{V}(z)\cos{\phi_\rho(z)}\cos{(\theta_\rho(z)+\Delta kz)}$ where $\Delta k = k_R-k_L$, $\mathcal{V}(z)=\tilde{\mathcal{V}}(z)\frac{2}{\pi\alpha}$ and a Gaussian disorder with impurity density $n_\mathrm{imp}$ and potential $\tilde{\mathcal{V}}(z) = \sum_i \mathcal{V}_0 \delta(z-z_i)$ is considered.

The 1D localization effect can be approached via the perturbative renormalization group analysis, e.g., for spinless\cite{RGimpurityspinless} and more complicated spinful\cite{RGimpurityspinful} cases. This is in fact starting from the delocalized phase and cannot go deep into the localized phase above the scale of the localization length $L$ since it will flow to strong coupling. On the other hand, directly inspecting the massive localized phase, variational method should prevail. Indeed, based on a charge-density wave picture\cite{SCHA1}, the phase field pinned to the impurities competes with its quantum fluctuations due to the 'elastic term' of that field in the bosonized Hamiltonian. A compromise is achieved when the phase field adjusts to the random potential over $L$, much longer than the average distance between impurities. Along this thought, we adopt the self-consistent harmonic approximation method\cite{SCHA_Boson1} that is similar to the more general variational theorem in path integral\cite{FeynmanStatMech}. Each field variable is decomposed to a classical part responsible for the compromised pinning and a quantum fluctuating part. The impurity effects enter via introducing variational mass terms dependent on $L$. This approach is good for very repulsive fermion interactions\cite{GiamarchiBook}, which is just suitable for our situation. 

Although due to the influence from other wires, the effective 1D model becomes involved and lacks a straightforward Hamiltonian form, it is still possible to evaluate the system's energy from the path integral. Variationally, we find\cite{SM}\mycomment{~\ref{App:Localization}} the localization length $L\propto {\mathcal{D}_\mathrm{imp}}^{\frac{1}{\eta-3}}$
where, as one would expect, only $\mathcal{D}_\mathrm{imp}=n_\mathrm{imp}\mathcal{V}_0^2$, which fully characterizes the Gaussian disorder, appears. 
In the localized regime $\eta<3$, $L$ remains finite. If $\eta$ could go beyond $3$, there would be a delocalization transition. However, this is not possible for our system with only Coulomb interactions where $\eta(v_\pm,v_g)$ decreases from $2$ upon increasing \mycomment{$\frac{v_g}{v^+}$}$v_g$ from $0$ to $\infty$ as shown in Fig.~\ref{fig:multiExponent}(b). 
As pointed out by previous studies\cite{RGimpurityspinless,RGimpurityspinful,SCHA_Boson2}, insulator-to-metal transition happens when there is increasingly attractive interaction where superconducting fluctuations predominates over disorder effects. In our Coulombic system, we therefore can only observe the enhanced localization effect, in agreement with another diagrammatic study\cite{localizationWeylSM} discussing a tendency to localization led by interaction and disorder.
It is worth addressing that although the TLL is induced by the field, the exponent $\gamma$ or $\eta$ does not depend on $B$, dissimilar to the quadratic band case where the Fermi velocity gains a $B$-dependence in the first place\cite{Glazman}. On one hand, in the low-energy regime, we end up with an effective 1D model with only the $B$-independent combination $Q^{*2}A_\perp$ present
. This is not surprising because the guiding center representation we introduced is in fact used at its continuum limit ($l_B\rightarrow0$)
. On the other hand, this means that, up to the leading order effect at the strong-field limit, the system behaves in a way independent to the field.

\textit{Temperature dependence of resistivity.---}In order to relate the system to the most common measurement technique, we study the temperature dependence of transport. Instead of calculating the conductivity that is inversely dependent on the scattering, a beneficial way is the memory function method\cite{Mori1,*MemoFunc,MemoFunc1D1,*MemoFunc1D4} and directly looking at the resistivity, which  corresponds to the diagramatic expansion taking into account both the vertex correction and the self-energy. 
Within the lowest order of coupling, a particular merit in practice is that one can evaluate the correlation function over the Hamiltonian free of disorder. 
By calculating an imaginary-time force-force correlation function $\mathcal{G}(\tau) = -\braket{\mathrm{T}_\tau F(z,\tau)F(z,0)}$ wherein the force operator $F=[j,H_\mathrm{imp}]$ with the current operator $j$, we find out\cite{SM}\mycomment{~\ref{App:resistivity}} that $\mathcal{G}(\tau) \propto \tau^{-\eta}$, whose Fourier transform $\mathcal{G}(\omega) \propto \tau^{1-\eta}$. Then when the typical energy scale is set by the temperature $\omega\sim T$, the memory function $\mathsf{M}(\omega) \propto \frac{\mathcal{G}(\omega)-\mathcal{G}(0)}{\omega} \propto \frac{\beta^{1-\eta}}{T} \propto \beta^{2-\eta}$ and hence we arrive at a resistivity's temperature dependence $\rho(T) \propto T^{\eta-2}$. 

There exists two energy (temperature) scales in this system\cite{RGimpurityspinful,NagaosaSingleImp}, the localization temperature $k_\mathrm{B}T_\mathrm{loc}=v_F/L$ and the discretization temperature $k_\mathrm{B}T_\mathrm{dis}=v_F n_\mathrm{imp}$, where we use $v_F$ to denote a typical Fermi velocity. Firstly, $T_\mathrm{dis}$ is the borderline of the correlation effect between impurities, above which, the single-impurity behavior prevails as a limiting resistivity $\rho(T\gg T_\mathrm{dis})\propto n_\mathrm{imp}$. For our TRB ($0<\eta<2$) or IB ($1<\eta<2$) system residing in the localized regime, as the temperature decreases, the resistivity will monotonically increase in contrast to the $\eta>2$ case where non-monotonic $\rho(T)$ could take place. Once the temperature traverses below $T_\mathrm{dis}$, the resistivity follows $\rho(T)\propto T^{\eta-2}$ for the dense Gaussian disorder situation, until the quantum interference from the disorder becomes more and more important when $T<T_\mathrm{loc}$, i.e., divergent $\rho(T) \propto T^{\eta-3}$ dominates at sufficiently low temperatures\cite{NagaosaSingleImp}. Note that $\eta<2$ due to the Coulomb interaction as shown in Fig.~\ref{fig:multiExponent}(b).

\textit{Large-$N$ behavior.---}Because of the point group symmetry in solids, experimentally realized Weyl semimetals usually possess many Weyl points. To bridge the gap between models and more realistic scenarios, we consider the situation comprising copies of our previous model. We use $N$ to count the pairs of opposite-chirality Weyl points in the first Brillouin zone. 
In the same spirit, we consider a similar bosonization problem of many quasi-1D wires with both intra-copy and inter-copy Coulomb interactions included, followed by deriving the 1D model of a single wire. 
For the impurity effects, following the previous formalism, we can take all the intra-copy impurity scatterings into account. 
By minimizing the total energy excess, we accordingly obtain an $N$-dependent exponent $\eta_N(v_\pm,v_g)$ 
that enters the temperature dependence of resistivity\cite{SM}\mycomment{~\ref{App:multi-pair}}. On the other hand, from the action expressed using the replica method\cite{replica}, we can make a Wilsonian analysis to develop the first-order renormalization group equation for the impurity strength $\mathcal{D}_\mathrm{imp}$. Both the intra-copy and inter-copy impurity scatterings lead to the same form $\frac{\dd \mathcal{D}_\mathrm{imp}(l)}{\dd l} = (3-\eta_N) \mathcal{D}_\mathrm{imp}(l)$. This means, taking all the impurity scatterings into account, the exponent will just be given by $\eta_N$.

It is important to note that all the previous conclusions on Green's function, localization and resistivity also hold for the TRB case with any $N$ while an even $N$ is for the IB case. Furthermore, as calculation shows, when $v_\pm=v$ and $N$ is the same, the two cases share the same $\gamma$ or $\eta$. 
Remarkabll, $\gamma$ is not directly related to $\eta$ in contrast to the simple formula $(K_\rho+K_\rho^{-1})/2$ in the standard case. This is because, due to the inter-wire Coulomb interactions, the effective action $\mathcal{S}_\mathrm{1D}$ contains rather complex momentum-frequency dependence [Eq.~(S43) of SM\cite{SM}]. 
As shown in Fig.~\ref{fig:multiExponent}, the multi-pair $\gamma_N$ ($\eta_N$) decreases (increases) with $N$. $\gamma_N$ increases from $1$ and diverges asymptotically proportional to $\sqrt{v_g/N}$ while $\eta_N$ ranges from $2$ to $2-2/N$ upon increasing $v_g$ from $0$ to $\infty$. This can be understood with the many-wire picture we rely on. When there are $N$ pairs, one has the freedom to place the corresponding guiding center lattices in the $x$-$y$ plane as uniform as possible to form sublattices of the original sparsest one. As aforementioned, the Coulomb interaction will be screened by the wires resultant from any copy. Therefore, the denser packing of the wires entails stronger screening and weaker interaction effects. Hence, the exponent $\eta_N$, although cannot exceed its noninteracting value $2$, approaches $2$ more quickly when $N$ increases. And when the Coulomb interaction is extremely strong, i.e., $v_g\rightarrow\infty$, the deviation, $2-\eta_N$, is exactly inversely proportional to $N$.

\textit{Estimation for TaP.---}Let us estimate the exponent of the Weyl semimetal TaP, which, known to date, might have the simplest stucture of Weyl points and is beneficial to revealing the interested physics\cite{TaP1,TaP2,predict3,predict4}. Among all the 12 pairs of Weyl points therein, 8 pairs well separated in momentum space off the $k_z=0$ plane are found to locate at the chemical potential while others lie rather lower, possibly leading to only 8 pairs determining the low-energy physics. We thus set $N=8$ and take the typical values of Fermi velocities and relative permittivity\cite{predict4,dielectric1,*dielectric2}, $v_R=2\times 10^5\mathrm{m/s},v_L=1\times 10^5\mathrm{m/s}$, $\varepsilon_r=10$ and hence $v_g=0.35\times 10^5\mathrm{m/s}$, and get $\eta_N=1.83$. 
Experimentally, to observe this, one should keep the temperature or frequency lower than the cyclotron gap to assure the dominance of the 1D channels.


\textit{Acknowledgments.---}
X.-X.Z thanks M. Ezawa and H. Ishizuka for helpful conversations at an early stage. X.-X.Z was supported by 
the Grant-in-Aid for JSPS Fellows (No.~16J07545). This work was supported by JSPS Grant-in-Aid for Scientific Research (No.~24224009) and JSPS Grant-in-Aid for Scientific Research on Innovative Areas (No.~26103006) from the Ministry of Education, Culture, Sports, Science and Technology (MEXT) of Japan and the ImPACT Program of Council for Science, Technology and Innovation (Cabinet office, Government of Japan). This work was also supported by CREST, Japan Science and Technology Agency.


%


\let\oldaddcontentsline\addcontentsline
\renewcommand{\addcontentsline}[3]{}
\let\addcontentsline\oldaddcontentsline


\newpage
\onecolumngrid
\newpage
{
	\center \bf \Large 
	Supplemental Material\\
	\large for ``Tomonaga-Luttinger liquid and localization in Weyl semimetals"\vspace*{0.1cm}\\ 
	\vspace*{0.0cm}
}
\begin{center}
	Xiao-Xiao Zhang$^1$ and Naoto Nagaosa$^{1,2}$\\
	\vspace*{0.15cm}
	\small{\textit{$^1$Department of Applied Physics, The University of Tokyo, 7-3-1 Hongo, Bunkyo-ku, Tokyo 113-8656, Japan
	\\$^2$RIKEN Center for Emergent Matter Science (CEMS), 2-1 Hirosawa, Wako, Saitama 351-0198, Japan
	}}\\
	\vspace*{0.25cm}	
	\end{center}
\twocolumngrid	

\tableofcontents

\setcounter{equation}{0}
\setcounter{figure}{0}
\setcounter{table}{0}
\setcounter{page}{1}
\renewcommand{\theequation}{S\arabic{equation}}
\renewcommand{\thefigure}{S\arabic{figure}}
\renewcommand{\bibnumfmt}[1]{[S#1]}
\renewcommand{\citenumfont}[1]{S#1}

\section{Weyl semimetal with Coulomb interaction}\label{App:models}
\subsection{A model Weyl semimetal under magnetic field}\label{App:WSMmodel}
As a preliminary, let us first turn on an external magnetic field $\vec{B}=B\hat{z}$ for a general Weyl Hamiltonian $\sum_i\epsilon_i\hbar k_i \sigma_i$, in which $\epsilon_i=\pm 1$ and we assume the velocity equal to unity for simplicity. To solve this, we resort to the commutation relation between gauge invariant mechanical momenta $[p_x,p_y]=-\ii\hbar eB$ derived from Peierls substitution $\hbar k_i\rightarrow p_i=-\ii \hbar\partial_i+eA_i\,,i=x,y$. Resembling to a harmonic oscillator, one can define $b=(p_x-\ii p_y)/E_B$ and $b^\dag=(p_x+\ii p_y)/E_B$ satisfying $[b,b^\dag]=1$ where $E_B=\sqrt{2\hbar eB}$. Then the two-by-two Hamiltonian can be easily solved, giving rise to eigenenergy $E_n=\pm\sqrt{n{E_B}^2+{E_z}^2}\,,n\geq 1$ with $E_z=\hbar k_z$. In addition to this, we get one more intriguing zero mode that is essential to the chiral anomaly, $E_0=-\chi E_z$, which doesn't shift with respect to the external magnetic field. The charge of the Weyl point is given by\cite{VolovikSM} $\chi=\mathrm{sgn}[\epsilon_1\epsilon_2\epsilon_3]$ and the pseudospin part of wavefunction reads $\ket{\downarrow}=(0,1)^\mathrm{T}$ or $\ket{\uparrow}=(1,0)^\mathrm{T}$ for $\epsilon_1\epsilon_2=\pm 1$, respectively. Also, the separation between eigenenergies scales as $\sqrt{B}$ instead of linear in $B$ for the quadractically dispersed electrons. The major consequence is that, in the presence of an external magnetic field, a 1D linearly dispersed mode along $\hat{z}$ direction is created, whose separation from other higher-energy eigenstates is larger than that of quadratic electrons. This suggests that, by turning on an adequate magnetic field, one can drive the Weyl fermion system to the quantum limit and the low-energy physical properties will depend mainly on the 1D mode singled out.

Let us exemplify with the noncentrosymmetric Weyl semimetal model presented in the main text, which has the mimimum of Weyl points according to the Nielson-Ninomiya theorem, it is straightforward to get the low-energy Weyl Hamiltonians of the form $\Delta\vec{k}\cdot\vec{\sigma}$ around the four Weyl points $(0,0,-k_R)\,,(0,0,-k_L)\,,(0,0,k_L)\,,(0,0,k_R)$ from left to right along the $k_z$ axis, whose $\vec{d}$ vectors read 
\begin{equation*}
\begin{split}
&(-\sin k_R \Delta k_x,\Delta k_y,\cos k_{RL}\sin k_R \Delta k_z)\\
&(-\sin k_L \Delta k_x,\Delta k_y,-\cos k_{RL}\sin k_L \Delta k_z)\\
&(\sin k_L \Delta k_x,\Delta k_y,\cos k_{RL}\sin k_L \Delta k_z)\\
&(\sin k_R \Delta k_x,\Delta k_y,-\cos k_{RL}\sin k_R \Delta k_z)
\end{split}
\end{equation*}
respectively, wherein $\cos k_{RL}$ is a shorthand for $\cos k_L-\cos k_R$ and these momenta are the deviations from the corresponding Weyl points. For any corresponding Weyl Hamiltonians $h_\kappa(\vec{k})$, the zero-mode eigenenergy and wavefunction under the magnetic field are 
\begin{equation}\label{eq:eigenenergy}
\varepsilon_\kappa(k_z)=rv_r(k_z-jk_r)
\end{equation} and 
\begin{equation}
\tilde{\varphi}_\kappa^{k_z}(z) = \frac{1}{\sqrt{\Omega}}\ee^{\ii(k_z-jk_r)z}\beta_{j}
\end{equation}
 wherein $\beta_{j=\mp 1}=\ket{\uparrow/\downarrow}$ and $\Omega$ is the system length along $\hat{z}$ direction.


\subsection{Coulomb interaction between quasi-1D wires}\label{App:Coulomb}
For the 1D linear modes singled out by an external magnetic field, we classify the possible scattering processes due to the Coulomb interaction. Consider a Coulombic two-body scattering $\braket{1,2\vert \hat{U} \vert 4,3}c_1^\dag c_2^\dag c_3c_4$ from electron states labeled as $4,3$ to $1,2$, one can find all the possible processes, for a fixed state $\kappa_1=(j_1,r_1)$ of electron $1$, by listing all the cases of $\epsilon_i^{j/r}=\pm\,,i=4,2,3$, which are defined by $j_1=\epsilon_4^j j_4=\epsilon_2^j j_2=\epsilon_3^j j_3$ and $r_1=\epsilon_4^r r_4=\epsilon_2^r r_2=\epsilon_3^r r_3$. This will include forward scatterings, backward scatterings and Umklapp scatterings at special fillings. It can also be reorganized to meet the current algebra classification used for Hubbard rung chains\cite{LadderChain1SM} except that we don't need to include spin-dependent scatterings in the current problem. The four types of forward scatterings in the main text can be denoted by $(\epsilon_4^j,\epsilon_2^j,\epsilon_3^j)$ and $(\epsilon_4^r,\epsilon_2^r,\epsilon_3^r)$
\begin{equation}\label{eq:4scattering}
\begin{split}
&1)\quad (+,+,+) \qquad (+,+,+) \\
&2)\quad (+,+,+) \qquad (+,-,-) \\
&3)\quad (+,-,-) \qquad (+,+,+) \\
&4)\quad (+,-,-) \qquad (+,-,-).
\end{split}
\end{equation}
In addition to this, we need to handle the Landau level degeneracy. By displacing the center of orbit $\vec{R}$, one can get other eigenstates of an annihilation operator
\begin{equation}\label{eq:CSbasis}
\chi_{\vec{R}}(\vec{r}_\perp)=\frac{1}{\sqrt{2\pi}l_B}\ee^{-[(\vec{r}_\perp-\vec{R})^2+2\ii\vec{r}_\perp\times\vec{R}]/4{l_B}^2}
\end{equation}
 where $\vec{r}_\perp$ is the coordinate in the $x$-$y$ plane. These are just the coherent states localized around $\vec{R}$. 
We can use the this basis to expand the electron field operator 
\begin{equation}\label{eq:Rexpand}
\Psi(\vec{r})=\sum_{\vec{R}}{\chi_{\vec{R}}(\vec{r}_\perp)\psi_{\vec{R}}(z)},
\end{equation} in which the on-wire electron field is expressed using four possible 1D modes
\begin{equation}\label{eq:psikappaexpand}
\psi_{\vec{R}}(z) = \sum_\kappa{\psi_{\kappa\vec{R}}(z)\beta_\kappa}.
\end{equation}
Similarly, one can also define
$\Psi(\vec{r})=\sum_\kappa{\Psi_\kappa(\vec{r})}$, in which the electron field of mode $\kappa$ is
\begin{equation}
\Psi_\kappa(\vec{r}) = \sum_{\vec{R}}{\chi_{\vec{R}}(\vec{r}_\perp)\psi_{\kappa\vec{R}}(z)\beta_\kappa}.
\end{equation}
Conventionally, the on-wire electron field of a particular mode $\kappa$ has its Fourier expansion $\psi_{\kappa\vec{R}}(z) = \frac{1}{\sqrt{\Omega}}\ee^{\ii k_zz}\psi_{\kappa\vec{R}}(k_z)$. For conciseness, here we stick to $\kappa$ to distinguish different 1D modes although the pseudospin wavefunction $\beta$ only depends on the side index $j$. An important aspect of the the coherent states is that they are not orthogonal, albeit over-complete. Instead, one can attain an asymptotic orthogonality 
$\braket{\chi_{\vec{R}}(\vec{r}_\perp)\vert\chi_{\vec{R}'}(\vec{r}_\perp)} \rightarrow 2\pi {l_B}^2 \delta^2(\vec{R}-\vec{R}')$ when $l_B\rightarrow 0$ while for a discrete lattice of $\vec{R}$ it becomes $\braket{\chi_{\vec{R}}(\vec{r}_\perp)\vert\chi_{\vec{R}'}(\vec{r}_\perp)} \rightarrow \delta_{\vec{R},\vec{R}'}$. We will use this relation to arrive at the quasi-1D bosonized Hamiltonian in the following. This corresponds to the limit of strong magnetic fields such that the magnetic length $l_B$ is much smaller than the characteristic $|\vec{R}-\vec{R}'|$, which should be valid for the long-range interaction we consider. In terms of this, the Coulomb potential admits an approximation $U(|\vec{r}-\vec{r}'|)=\frac{e^2}{|\vec{r}-\vec{r}'|}=\frac{e^2}{\sqrt{(z-z')^2+(\vec{r}_\perp-\vec{r}_\perp')^2}} \approx \frac{e^2}{\sqrt{(z-z')^2+(\vec{R}-\vec{R}')^2}}$, where $\vec{r}_\perp$ is replaced by the guiding center $\vec{R}$ since the deviation away from the wire is negligibly small due to the transverse confinement. We will refer to this as $U(z_-,R_-)$ wherein $z_-=z-z'\,,\vec{R}_-=\vec{R}-\vec{R}'$.

From Eq.~\eqref{eq:eigenenergy}, the noninteracting Hamiltonian for all the 1D modes is given by 
\begin{equation}\label{eq:H_0ver1}
\begin{split}
H_0 &= \sum_{\kappa\vec{R}}{\int\dd z {\psi_{\kappa\vec{R}}^\dag(z) \varepsilon_\kappa(-\ii\partial_z) \psi_{\kappa\vec{R}}(z)}} \\
&= \sum_{\kappa k_z\vec{R}}{\varepsilon_\kappa(k_z) \psi_{\kappa\vec{R}}^\dag(k_z) \psi_{\kappa\vec{R}}(k_z)}.
\end{split}
\end{equation}
With Eq.~\eqref{eq:Rexpand} and the asymptotic orthogonality, the Coulomb interaction part of the Hamiltonian takes the form (up to unimportant chemical potential terms)
\begin{equation}
\begin{split}
H_I &= \frac{1}{2} \int\dd\vec{r}\dd\vec{r}' U(|\vec{r}-\vec{r}'|) \Psi^\dag(\vec{r})\Psi(\vec{r}) \Psi^\dag(\vec{r}')\Psi(\vec{r}') \\
&= \frac{1}{2} \sum_{\vec{R}_1\vec{R}_2\vec{R}_3\vec{R}_4} \int\dd z\dd z' U(|\vec{r}-\vec{r}'|) \braket{\chi_{\vec{R}_1}(\vec{r}_\perp)\vert\chi_{\vec{R}_4}(\vec{r}_\perp)} \\
&\times\braket{\chi_{\vec{R}_2}(\vec{r}_\perp')\vert\chi_{\vec{R}_3}(\vec{r}_\perp')} \psi_{\vec{R}_1}^\dag(z)\psi_{\vec{R}_4}(z)\psi_{\vec{R}_2}^\dag(z')\psi_{\vec{R}_3}(z') \\
&= \frac{1}{2} \sum_{\vec{R}\vec{R}'} \int\dd z\dd z' U(z_-,R_-)  \psi_{\vec{R}}^\dag(z)\psi_{\vec{R}}(z)\psi_{\vec{R}'}^\dag(z')\psi_{\vec{R}'}(z').
\end{split}
\end{equation}
Then, using Eq.~\eqref{eq:psikappaexpand}, one can further reduce it to 
\begin{equation}\label{eq:H_intver1}
\begin{split}
H_I &= \frac{1}{2} \sum_{\vec{R}\vec{R}'} \int\dd z\dd z' U(z_-,R_-)  \sum_{\kappa_1\kappa_2\kappa_3\kappa_4} \braket{\beta_{\kappa_1}\vert\beta_{\kappa_4}} \\
&\times\braket{\beta_{\kappa_2}\vert\beta_{\kappa_3}} \psi_{\kappa_1\vec{R}}^\dag(z)\psi_{\kappa_4\vec{R}}(z)\psi_{\kappa_2\vec{R}'}^\dag(z')\psi_{\kappa_3\vec{R}'}(z') \\
&= \frac{1}{2} \sum_{\vec{R}\vec{R}'} \int\dd z\dd z' U(z_-,R_-) \\
&\times\sum_{\kappa\kappa'} \psi_{\kappa\vec{R}}^\dag(z)\psi_{\kappa\vec{R}}(z)\psi_{\kappa'\vec{R}'}^\dag(z')\psi_{\kappa'\vec{R}'}(z'),
\end{split}
\end{equation}
wherein the inner products between pseudospin states are evaluated for the four types of scatterings (Eq.~\eqref{eq:4scattering}) we mentioned.

\subsection{Bosonization}\label{App:bosonization}
Now we are ready to study the system of many interacting quasi-1D wires through the bosonization method. 
As noted in the main text, in order to handle the unequal chiral velocities, we make use of the more original chiral boson field $\varphi^\kappa=\varphi^{j,r}$ that bosonizes a Weyl fermion in $(1+1)$-dimensions. 
One has the commutation relation for the same wire
\begin{equation}\label{eq:chiralbosonCommutation}
[\nabla\varphi^\kappa(z),\varphi^\kappa(z')]=\ii 2\pi r\delta(z-z')
\end{equation}
and the electron density
\begin{equation}\label{eq:chiralbosonRho}
\rho^\kappa=\frac{1}{2\pi}\nabla\varphi^\kappa,
\end{equation}
in which the wire index $\vec{R}$ is omitted.

Henceforth, as suggested by the aforementioned strong-field limit $l_B \rightarrow 0$, we will rely on the continuum expressions for the guiding center lattice $\vec{R}$, i.e., $\sum_{\vec{R}}=\frac{1}{A_\perp}\int\dd\vec{R}$. 
Then the bosonized form of Eq.~\eqref{eq:H_0ver1} becomes 
\begin{equation}\label{eq:H_0ver2}
\begin{split}
H_0 &= \frac{1}{4\pi}\sum_{\kappa\vec{R}}\int\dd z v_r (\nabla\varphi_{\vec{R}}^\kappa)^2 \\
&= \frac{1}{4\pi A_\perp}\sum_{\kappa}\int\dd \vec{r} v_r (\nabla\varphi_{\vec{R}}^\kappa)^2 \\
&= \frac{1}{4\pi A_\perp V}\sum_{\kappa\vec{k}} v_r q^2\varphi_{\vec{k}}^\kappa\varphi_{-\vec{k}}^\kappa
\end{split}
\end{equation}
where the gradient operator only applies to coordinate $z$ henceforth.
On the other hand, the interaction part (Eq.~\eqref{eq:H_intver1}) can be divided to four parts $H_I=\sum_iH_i$ with
\begin{equation}
\begin{split}
H_i &= \frac{1}{2}\sum_{\kappa\vec{R}\vec{R}'}\int\dd z\dd z' U(z_-,R_-) \rho_{\vec{R}}^\kappa(z)\rho_{\vec{R}'}^{\kappa_i}(\vec{r}') \\
&= \frac{1}{2}\frac{1}{(2\pi)^2}\sum_\kappa\frac{1}{A_\perp^2}\int\dd\vec{r}\dd\vec{r}' U(z_-,R_-) \nabla\varphi_{\vec{R}}^\kappa(z)\nabla\varphi_{\vec{R}'}^{\kappa_i}(z') \\
&= \frac{1}{2}\frac{1}{(2\pi)^2}\sum_\kappa\frac{1}{A_\perp^2\pi}\sum_{q,\vec{Q}}\frac{4\pi e^2 q^2}{q^2+Q^2} \nabla\varphi_{\vec{k}}^\kappa\nabla\varphi_{-\vec{k}}^{\kappa_i} \\
&= \frac{g}{4\pi A_\perp V}\sum_{\kappa\vec{k}}\frac{q^2}{k^2} \nabla\varphi_{\vec{k}}^\kappa\nabla\varphi_{-\vec{k}}^{\kappa_i},
\end{split}
\end{equation}
in which $g=\frac{2e^2}{A_\perp}$ and Eq.~\eqref{eq:chiralbosonRho} and the Fourier transform of the Coulomb potential are used in the second and the third equalities, respectively.
The index $i$ signifies the $i$th scattering in Eq.~\eqref{eq:4scattering} and accordingly $\kappa_i=(j,r)\,,(j,-r)\,,(-j,r)\,,(-j,-r)$ for $i=1,2,3,4$ respectively.
Using Eq.~\eqref{eq:chiralbosonCommutation}, we are ready to write down the action of this system in Euclidean spacetime
\begin{equation}\label{eq:chiralbosonAction}
\begin{split}
\mathcal{S} &= \sum_\kappa \int_0^\beta\dd\tau\dd\vec{r}\frac{-\ii r}{4\pi} \varphi^\kappa(\vec{r},\tau)\partial_\tau\nabla\varphi^\kappa(\vec{r},\tau) + \int_0^\beta\dd\tau H \\
&= \frac{1}{4\pi\beta A_\perp V} \sum_{\kappa, p} \lbrace -\ii rq\omega\varphi^\kappa_{-p}\varphi^\kappa_{p} \\
&+ q^2 [ (v_r+\frac{g}{k^2})\varphi^\kappa_{-p}\varphi^\kappa_{p} + \frac{g}{k^2} \sum_{i=2,3,4} \varphi^\kappa_{-p}\varphi^{\kappa_i}_{p} ] \rbrace \\
&= \frac{1}{4\pi\beta A_\perp V} \sum_p \vec{\varphi}^\dag_p W_p \vec{\varphi}_p
\end{split}
\end{equation}
wherein $\vec{\varphi}=(\varphi^{\kappa_1},\cdots,\varphi^{\kappa_4})^\mathrm{T}$ and the Fourier expansion is defined as $\varphi(\vec{r},\tau)=(\beta V)^{-\frac{1}{2}}\sum_{\vec{k},\omega}\varphi_p\ee^{\ii(\vec{k}\cdot\vec{r}-\omega\tau)}$. The action matrix 
\begin{equation}
W_p = q^2
\begin{bmatrix}
V_R-\frac{\zz}{q} & V_g & V_g & V_g \\
V_g & V_L+\frac{\zz}{q} & V_g & V_g \\
V_g & V_g & V_R-\frac{\zz}{q} & V_g \\
V_g & V_g & V_g & V_L+\frac{\zz}{q}
\end{bmatrix}
\end{equation} with $V_g(k)=\frac{g}{k^2}$, $V_r=v_r+V_g$ and $\zz$, which can equal $\ii\omega$ for instance, is a generic complex frequency not to be confused with the coordinate $z$ in real space. Note that we have used the fact that $\varphi(\vec{r},\tau)$ is real. This can be block-diagonalized by transforming to new fields 
\begin{equation}\label{eq:jbosontransform1}
\vec{\xi}=(\theta^R,\theta^L,\phi^R,\phi^L)^\mathrm{T}=\frac{1}{2}
\begin{bmatrix}
I & -I \\
-I & -I 
\end{bmatrix}
\vec{\varphi},
\end{equation}
upon which, the action becomes
\begin{equation}\label{eq:jbosonAction}
\begin{split}
\mathcal{S} &= \frac{1}{2\pi\beta A_\perp V} \sum_p q^2 \left[[\theta^R,\theta^L]_{-p}
\begin{bmatrix}
v_R - \frac{\zz}{q} & 0 \\
0 & v_L+\frac{\zz}{q} 
\end{bmatrix}
\begin{bmatrix}
\theta^R \\
\theta^L 
\end{bmatrix}_p \right. \\
&\left. 
+ [\phi^R,\phi^L]_{-p} 
\begin{bmatrix}
2V_g+v_R - \frac{\zz}{q} & 2V_g \\
2V_g & 2V_g+v_L+\frac{\zz}{q} 
\end{bmatrix}
\begin{bmatrix}
\phi^R \\
\phi^L 
\end{bmatrix}_p
\vphantom{\begin{bmatrix}
\theta^R \\
\theta^L 
\end{bmatrix}_p}
\right].
\end{split}
\end{equation}
On the other hand, for the TRB case, working in the standard $(\phi,\theta)$ fields with $\varphi^r=-(\phi-r\theta)$, we have
\begin{equation}\label{eq:TRBAction}
\begin{split}
\mathcal{S} = \frac{1}{2\pi\beta A_\perp V} \sum_p q^2 
[\phi,\theta]_{-p} 
\begin{bmatrix}
2V_g+v & \zz/q \\
\zz/q & v
\end{bmatrix}
\begin{bmatrix}
\phi \\
\theta 
\end{bmatrix}_p.
\end{split}
\end{equation}
Transformation \eqref{eq:jbosontransform1} is nothing but combining fields of different side index $j$
\begin{equation}\label{eq:jbosontransform2}
\theta^r = \frac{1}{2}(\varphi^{1,r}-\varphi^{-1,r})\,,\phi^r = -\frac{1}{2}(\varphi^{1,r}+\varphi^{-1,r}),
\end{equation}
which is useful in the current problem since 1D modes with only different side index $j$ share the same velocity in much the same way as chirality index $r$ does in the standard case where opposite-chirality fields are combined. 
The new commutation relations read
\begin{equation}\label{eq:jbosonCommutation}
[\nabla\theta^r(z),\theta^r(z')] = [\nabla\phi^r(z),\phi^r(z')] = \ii\pi r\delta(z-z').
\end{equation}
It might as well be worth noting that, compared with the more standard action of the simplest bosonization case, the Berry phase term in Eq.~\eqref{eq:chiralbosonAction} or Eq.~\eqref{eq:jbosonAction} appears in the diagonal and is half of the value expected from the corresponding commutation rule Eq.~\eqref{eq:chiralbosonCommutation} or Eq.~\eqref{eq:jbosonCommutation}. This is because a chiral boson field and its spatial derivative are not independent.

\subsection{Green's function}\label{App:GreenFunc}
In this subsection, let us calculate the Green's functions of the chiral electrons. Feeding the new bosonic fields given in Eq.~\eqref{eq:jbosontransform2}, we have the electron Green's function of 1D mode $\kappa$ 
\begin{equation}\label{eq:electronGreen1}
\begin{split}
& -\mathcal{G}_\kappa(r) = \braket{\mathrm{T}_\tau \psi_\kappa(r)\psi_\kappa^\dag(0)} \\
&= \frac{\ee^{\ii jk_rz}}{2\pi\alpha} \braket{\mathrm{T}_\tau \ee^{\ii r(j\theta^r-\phi^r)(r)} \ee^{-\ii r(j\theta^r-\phi^r)(0)}} \\
&= \frac{\ee^{\ii jk_rz}}{2\pi\alpha} \ee^{-\frac{1}{2}\braket{(\theta^r(r)-\theta^r(0))^2+ (\phi^r(r)-\phi^r(0))^2}} \ee^{\pm \ii \pi \Theta(-\tau)},
\end{split}
\end{equation}
where we use the Debye-Waller formula for quadratic action and the fact that $\theta$ and $\phi$ fields are decoupled. The exponential with a Heaviside step function $\Theta(-\tau)$ can be dropped since we will simply focus on the $\tau>0$ case. In addition, we only consider the correlation on a particular wire and set $\vec{R}=0$ without loss of generality, i.e., $r=(z,\vec{0},\tau)$.
After lengthy calculations presented in Sec.~\ref{App:generalGreen}, we obtain the electron Green's function
\begin{equation}
\begin{split}
& -\mathcal{G}_\kappa(z,\tau) = \braket{\mathrm{T}_\tau \psi_\kappa(r)\psi_\kappa^\dag(0)} 
= \frac{\ee^{\ii jk_rz}}{2\pi\alpha} \left[ \frac{\alpha+v_r\tau-\ii r z}{\alpha} \right]^{\frac{-1}{2}} \\
& \times \prod_{\lambda=\pm} \left[ \frac{\alpha+\frac{\tau}{2}(w^*+\lambda v^-)  -\ii r z}{\alpha} \right]^{-\frac{1}{2} \frac{4V_g^*+v^++\lambda rw^*}{2w^*}}
\end{split}
\end{equation}
where $w^*=\sqrt{v^+(8V_g^*+v^+)}\,,V_g^*=\frac{g}{{Q^*}^2}$. At the long-distance ($z\gg\alpha$) limit, $\mathcal{G}_\kappa(z,0) \sim z^{-\gamma}$ where $\gamma=\frac{1}{2} \left( 1+ \sum_{\lambda=\pm} \frac{4V_g^*+v^++\lambda rw^*}{2w^*} \right) = \frac{1}{2} \left( 1+ \frac{4V_g^*+v^+}{w^*} \right)>1$. On the other hand, if one only wants to extract the correct $\gamma$ value instead of some concrete form of the general Green's function, one can turn to another approach sketched in Sec.~\ref{App:asympGreen}.

\subsubsection{General form}\label{App:generalGreen}
We can invert the action Eq.~\eqref{eq:jbosonAction} to get the propagators
\begin{equation}\label{eq:jbosonCorr_pspace}
\begin{split}
\braket{\theta^r_{-p}\theta^r_{p}} &= \pi\beta A_\perp V\frac{1}{-rq(\zz-\zz_{0r})} \\ 
\braket{\phi^r_{-p}\phi^r_{p}} &= \pi\beta A_\perp V \frac{q(2V_g+v_{-r})+\zz r}{-q(\zz-\zz_+)(\zz-\zz_-)},
\end{split}
\end{equation}
where $\zz_{0r}=rqv_r$ and $\zz_{\pm}=\frac{q}{2}(v^-\pm\sqrt{v^+(8V_g+v^+)})$ with $v^\pm = v_R \pm v_L>0$. Separating the prefactors, residues at the three poles are $\mathrm{Res}_\lambda^r=\frac{-1}{rq}$ and $\mathrm{Res}_\lambda^r=\frac{4V_g+v^+ + \lambda r \sqrt{v^+(8V_g+v^+)}}{-2\lambda q\sqrt{v^+(8V_g+v^+)}}$ for $\lambda=0$ and $\lambda=\pm$, respectively. The correlation functions
\begin{subequations}\label{eq:jbosonCorrelations}
\begin{align}\label{eq:jbosonCorrTheta}
\braket{(\theta^r(r)-\theta^r(0))^2} = \frac{1}{(\beta V)^2} \sum_{\vec{k}\omega_n} \braket{\theta^r_{-p}\theta^r_{p}} A(r\cdot p)
\end{align}
\begin{align}\label{eq:jbosonCorrPhi}
\braket{(\phi^r(r)-\phi^r(0))^2} = \frac{1}{(\beta V)^2} \sum_{\vec{k}\omega_n} \braket{\phi^r_{-p}\phi^r_{p}} A(r\cdot p)
\end{align}
\end{subequations}
wherein $A=A(r\cdot p)=2-2\cos(qz-\omega_n\tau)=2-2\cos(qz+\ii\zz\tau)$.
For Eq.~\eqref{eq:jbosonCorrTheta}, we can perform the summation over bosonic matsubara frequencies as follows.
\begin{equation*}
\begin{split}
\sum_{q\omega_n}\frac{1}{-rq(\zz-\zz_{0r})} \ee^{\eta\zz} &= -\beta\sum_q \mathrm{Res}_0^r n_B(\zz_{0r}) \\
&= -\beta\sum_{q>0} \mathrm{Res}_0^r (1+2n_B(\zz_{0r})),
\end{split}
\end{equation*}
where $\zz$ is understood as $\ii\omega_n$, $n_B$ is bosonic distribution and we take $\eta\rightarrow 0^+$, but the result remains the same for $\eta\rightarrow 0^-$. To control the convergence at $\Re\zz<0$ and $\Re\zz>0$ we use bosonic weighting functions, $(1+n_B(\zz))$ and $n_B(\zz)$, respectively. Then we have
\begin{equation*}
\begin{split}
&\sum_{q\omega_n}\frac{1}{-rq(\zz-\zz_{0r})} \ee^{\ii qz-\tau\zz} \\
=& -\beta\sum_q \mathrm{Res}_0^r (1+n_B(\zz_{0r})) \ee^{\ii qz-\tau\zz_{0r}} \\
=& -\beta\sum_{q>0} \mathrm{Res}_0^r [ (1+n_B(\zz_{0r})) \ee^{\ii qz-\tau\zz_{0r}} + n_B(\zz_{0r}) \ee^{-\ii qz+\tau\zz_{0r}} ]
\end{split}
\end{equation*}
\begin{equation*}
\begin{split}
&\sum_{q\omega_n}\frac{1}{-rq(\zz-\zz_{0r})} \ee^{-\ii qz+\tau\zz} \\
=& -\beta\sum_q \mathrm{Res}_0^r n_B(\zz_{0r}) \ee^{-\ii qz+\tau\zz_{0r}} \\
=& -\beta\sum_{q>0} \mathrm{Res}_0^r [ n_B(\zz_{0r}) \ee^{-\ii qz+\tau\zz_{0r}} + (1+n_B(\zz_{0r})) \ee^{\ii qz-\tau\zz_{0r}} ]
\end{split}
\end{equation*}
And we arrive at 
\begin{equation}\label{eq:jbosonCorrTheta_pre}
\begin{split}
&\sum_{q\omega_n}\frac{1}{-rq(\zz-\zz_{0r})} A(r\cdot p) \\
=& -2\beta\sum_{q>0} \mathrm{Res}_0^r [ n_B(\zz_{0r})A_{\zz_{0r}} +1-\ee^{\ii qz-\tau\zz_{0r}} ] \\
\LongArrow{\beta\rightarrow\infty} & -2\beta\sum_{q>0} \mathrm{Res}_0^r r(1-\ee^{r(\ii qz-\tau\zz_{0r})} ),
\end{split}
\end{equation}
where $A_{\zz_{0r}} = 2-2\cos(qz+\ii\zz_{0r}\tau)$. 
Similarly, for Eq.~\eqref{eq:jbosonCorrPhi}, we have (summing only over $\lambda=\pm$)
\begin{equation*}
\begin{split}
\sum_{q\omega_n} \frac{q(2V_g+v_{-r})+\zz r}{-q(\zz-\zz_+)(\zz-\zz_-)} \ee^{\eta\zz} &= -\beta\sum_{q,\lambda} \mathrm{Res}_\lambda^r n_B(\zz_\lambda) \\
&= -\beta\sum_{q>0,\lambda} \mathrm{Res}_\lambda^r (1+2n_B(\zz_\lambda)).
\end{split}
\end{equation*}
\begin{equation*}
\begin{split}
&\sum_{q\omega_n} \frac{q(2V_g+v_{-r})+\zz r}{-q(\zz-\zz_+)(\zz-\zz_-)} \ee^{\ii qz-\tau\zz_\lambda} \\
=& -\beta\sum_{q\lambda} \mathrm{Res}_\lambda^r (1+n_B(\zz_\lambda)) \ee^{\ii qz-\tau\zz_\lambda} \\
=& -\beta\sum_{q>0,\lambda} \mathrm{Res}_\lambda^r [ (1+n_B(\zz_\lambda)) \ee^{\ii qz-\tau\zz_\lambda} + n_B(\zz_\lambda) \ee^{-\ii qz+\tau\zz_\lambda} ]
\end{split}
\end{equation*}
\begin{equation*}
\begin{split}
&\sum_{q\omega_n} \frac{q(2V_g+v_{-r})+\zz r}{-q(\zz-\zz_+)(\zz-\zz_-)} \ee^{-\ii qz+\tau\zz_\lambda} \\
=& -\beta\sum_{q\lambda} \mathrm{Res}_\lambda^r n_B(\zz_\lambda) \ee^{-\ii qz+\tau\zz_\lambda} \\
=& -\beta\sum_{q>0,\lambda} \mathrm{Res}_\lambda^r [ n_B(\zz_\lambda) \ee^{-\ii qz+\tau\zz_\lambda} + (1+n_B(\zz_\lambda)) \ee^{\ii qz-\tau\zz_\lambda} ]
\end{split}
\end{equation*}
And we arrive at 
\begin{equation}\label{eq:jbosonCorrPhi_pre}
\begin{split}
&\sum_{q\omega_n} \frac{q(2V_g+v_{-r})+\zz r}{-q(\zz-\zz_+)(\zz-\zz_-)} A(r\cdot p) \\
=& -2\beta\sum_{q>0,\lambda} \mathrm{Res}_\lambda^r [ n_B(\zz_\lambda)A_{\zz_\lambda} +1-\ee^{\ii qz-\tau\zz_\lambda} ] \\
\LongArrow{\beta\rightarrow\infty} & -2\beta\sum_{q>0,\lambda} \mathrm{Res}_\lambda^r \lambda(1-\ee^{\lambda(\ii qz-\tau\zz_\lambda)} ).
\end{split}
\end{equation}
Combining Eq.~\eqref{eq:jbosonCorr_pspace}, Eq.~\eqref{eq:jbosonCorrTheta} and Eq.~\eqref{eq:jbosonCorrTheta_pre}, we obtain the correlation function of the $\theta$ field
\begin{equation}
\begin{split}
&\braket{(\theta^r(r)-\theta^r(0))^2} \\
&= \frac{-2\pi}{\Omega} \sum_{q>0} \mathrm{Res}_0^r(n_B(\zz_{0r})A_{\zz_{0r}} +1-\ee^{\ii qz-\tau\zz_{0r}}) \\
&\LongArrow{\beta\rightarrow\infty} \frac{-2\pi}{\Omega} \sum_{q>0} \mathrm{Res}_0^r r (1-\ee^{r(\ii qz-\tau\zz_{0r})}),
\end{split}
\end{equation}
in which summation $\sum_{\vec{Q}} 1= \frac{\Omega_\perp}{A_\perp}$ is used. Combining Eq.~\eqref{eq:jbosonCorr_pspace}, Eq.~\eqref{eq:jbosonCorrPhi} and Eq.~\eqref{eq:jbosonCorrPhi_pre}, we obtain the correlation function of the $\phi$ field
\begin{equation}
\begin{split}
&\braket{(\phi^r(r)-\phi^r(0))^2} \\
&= \frac{-2\pi}{V} \sum_{\vec{Q}, q>0,\lambda} \mathrm{Res}_\lambda^r [ n_B(\zz_\lambda)A_{\zz_\lambda} +1-\ee^{\ii qz-\tau\zz_\lambda} ] \\
&\LongArrow{\beta\rightarrow\infty} \frac{-2\pi}{V} \sum_{\vec{Q}, q>0,\lambda} \mathrm{Res}_\lambda^r \lambda( 1-\ee^{\lambda(\ii qz-\tau\zz_\lambda)} ).
\end{split}
\end{equation}
At the zero temperature limit ($\beta\rightarrow \infty$), it is possible to proceed  by turning the momentum summation to integral with a lattice cutoff factor $\ee^{-\alpha q}$. Thus, 
\begin{equation}\label{eq:jbosonCorrTheta_final}
\begin{split}
\braket{(\theta^r(r)-\theta^r(0))^2} 
&=  \int_0^\infty \dd q \ee^{-\alpha q} \frac{1-\ee^{r(\ii qz-\tau\zz_{0r})}}{q} \\
&= \ln\frac{\alpha+v_r\tau-\ii r z}{\alpha},
\end{split}
\end{equation}
which is valid since $\alpha+v_r\tau>0$.
On the other hand, for the much more complex $\phi$ correlation, it is necessary to resort to some approximations. The $\vec{Q}$-integral is not within the range of analytic solution. Concerning the low-energy property of this system, the typical value of momentum $q$ should be negligibly small than the momentum $\vec{Q}$ of the guiding center lattice. Hence we replace $V_g=\frac{g}{k^2}$ by $V_g^*=\frac{g}{{Q^*}^2}$ for the nonce so as to relieve us of the $\vec{Q}$-integral, where $Q^*$ is some characteristic value of the momentum. This enables us to perform the $q$-integral
\begin{equation}\label{eq:jbosonCorrPhi_final}
\begin{split}
&\braket{(\phi^r(r)-\phi^r(0))^2} \\
&=  -\sum_\lambda \int_0^\infty \dd q \mathrm{Res}_\lambda^r \lambda( 1-\ee^{\lambda(\ii qz-\tau\zz_\lambda)} ) \ee^{-\alpha q} \\
&= \sum_\lambda \frac{4V_g^*+v^++\lambda rw^*}{2w^*} \ln \frac{\alpha+\frac{\tau}{2}(w^*+\lambda v^-)  -\ii r z}{\alpha},
\end{split}
\end{equation}
which is valid since $\alpha+\frac{\tau}{2}(w^*+\lambda v^-)>0$ and we define $w^*=\sqrt{v^+(8V_g^*+v^+)}$.
Finally, combining Eq.~\eqref{eq:electronGreen1}, Eq.~\eqref{eq:jbosonCorrTheta_final} and Eq.~\eqref{eq:jbosonCorrPhi_final}, we arrive at the electron Green's function written in the beginning.

\subsubsection{Asymptotic form}\label{App:asympGreen}
For the asymptotic behavior at long distance of the on-wire Green's functions, it is convenient to rely on the effective 1D model in Sec.~\ref{App:Eff1Dmodel}. Combining Eq.~\eqref{eq:ChargeChiTransform1} and Eq.~\eqref{eq:electronGreen1}, one has 
\begin{equation}\label{eq:electronAsympGreenIB}
\begin{split}
& \mathcal{G}_\kappa(z) \propto \exp \left[ -\frac{1}{2}\left( \frac{1}{4}\braket{[(\theta_\rho+r\theta_\chi)(z)-(\theta_\rho+r\theta_\chi)(0))]^2} \right.\right.\\
&\left.\left.+ \frac{1}{4}\braket{[(\phi_\rho+r\phi_\chi)(z)-(\phi_\rho+r\phi_\chi)(0))]^2} \vphantom{\frac{1}{2}}\right) \right].
\end{split}
\end{equation}
As shown above, the $\theta$ part gives a trivial exponent $\frac{1}{2}$. We thus simply apply the method in Sec.~\ref{App:resistivity} to the $\phi$ part
\begin{equation}\label{eq:electronAsympGreenIB1}
\begin{split}
& \exp \left[ -\frac{1}{2} \frac{1}{4}\braket{[(\phi_\rho(z)-\phi_\rho(0))+r(\phi_\chi(r)-\phi_\chi(0))]^2} \right] \\
&=\exp \left[ -\frac{1}{2} \frac{1}{4}\left(\braket{(\phi_\rho(z)-\phi_\rho(0))^2 + (\phi_\chi(z)-\phi_\chi(0))^2} \right.\right.\\
&\left. \left. + 2r\braket{(\phi_\rho(z)-\phi_\rho(0))(\phi_\chi(z)-\phi_\chi(0))} \right) \vphantom{\frac{1}{2}} \right].
\end{split}
\end{equation}
Previewing the notation in Sec.~\ref{App:solveSCeq}, the result is $\gamma=\frac{1}{2} + \frac{1}{4\pi}\int_0^{2\pi}{\dd\mu \frac{B-C'}{B^2\cos\mu}}$ or $\gamma=\frac{1}{2} + \frac{1}{4\pi}\int_0^{2\pi}{\dd\mu \frac{A-C'}{A^2\cos\mu}}$ for $r=\mp1$ where $C'=2\frac{v_{g'}}{v^+}\cos\mu$. In fact, $\gamma$ is independent to $\kappa$ and increases with $v_g$ from unity.
Similarly, for the TRB case, we have 
\begin{equation}\label{eq:electronAsympGreenTRB}
\begin{split}
& \mathcal{G}_r(z) \propto \braket{\ee^{-\ii(r\phi-\theta)(z)}\ee^{\ii(r\phi-\theta)(0)}} \\
&=\exp \left[ -\frac{1}{2} \left(\braket{(\phi(z)-\phi(0))^2 + (\theta(z)-\theta(0))^2} \right.\right.\\
&\left. \left. - 2r\braket{(\phi(z)-\phi(0))(\theta(z)-\theta(0))} \right) \vphantom{\frac{1}{2}} \right],
\end{split}
\end{equation}
which also gives a $\gamma>1$ increasing with $v_g$.

\section{Impurity effect on an effective 1D wire}\label{App:ImpurityIn1D}
\subsection{Effective 1D Luttinger liquid system}\label{App:Eff1Dmodel}
The charge-chirality separated basis can be attained by combining the opposite-chirality fields in Eq.~\eqref{eq:jbosontransform1} or Eq.~\eqref{eq:jbosontransform2}
\begin{equation}\label{eq:ChargeChiTransform1}
\begin{split}
\vec{\zeta} = (\theta_\rho,\theta_\chi,\phi_\rho,\phi_\chi)^\mathrm{T}
=
\begin{bmatrix}
H & 0 \\
0 & H 
\end{bmatrix}
\vec{\xi}
=\frac{1}{2}
\begin{bmatrix}
H & -H \\
-H & -H 
\end{bmatrix}
\vec{\varphi},
\end{split}
\end{equation}
wherein $H = 
\begin{bmatrix}
1 & 1 \\
1 & -1 
\end{bmatrix}$ is the order-$2$ Hadamard matrix. 
The new commutation relations read 
\begin{equation}\label{eq:ChargeChiCommutation}
[\nabla\theta_\rho(z),\theta_\chi(z')] = [\nabla\phi_\rho(z),\phi_\chi(z')] = \ii2\pi \delta(z-z').
\end{equation}
The total particle density of all modes is given by
\begin{equation}\label{eq:density}
\rho = -\frac{1}{\pi} \nabla\phi_\rho,
\end{equation}
which can be easily seen from Eq.~\eqref{eq:chiralbosonRho}. Upon the new fields $\vec{\zeta}$, the action of the system, Eq.~\eqref{eq:chiralbosonAction} or Eq.~\eqref{eq:jbosonAction}, is transformed into
\begin{equation}\label{eq:ChargeChiAction}
\begin{split}
\mathcal{S} &= \frac{1}{2\pi\beta A_\perp V} \sum_p \vec{\zeta}^\dag_p M_p \vec{\zeta}_p \\
&= \frac{1}{2\pi\beta A_\perp V} \sum_p \frac{q^2}{4} \left[[\theta_\rho,\theta_\chi]_{-p}
\begin{bmatrix}
v^+ & v^--\frac{2\zz}{q} \\
v^--\frac{2\zz}{q} & v^+ 
\end{bmatrix}
\begin{bmatrix}
\theta_\rho \\
\theta_\chi 
\end{bmatrix}_p \right. \\
&\left. 
+ [\phi_\rho,\phi_\chi]_{-p} 
\begin{bmatrix}
8V_g+v^+ & v^--\frac{2\zz}{q} \\
v^--\frac{2\zz}{q} & v^+ 
\end{bmatrix}
\begin{bmatrix}
\phi_\rho \\
\phi_\chi 
\end{bmatrix}_p
\vphantom{\begin{bmatrix}
\theta_\rho \\
\theta_\chi 
\end{bmatrix}_p}
\right]
\end{split}
\end{equation}
as shown in the main text.
%
%

Now we are ready to derive the effectice 1D model of the system. This can be done, without loss of generality, by integrating out the fields except the ones $\vec{\zeta}_0(z,\tau)=\vec{\zeta}(z,\vec{R},\tau)$ on a particular wire at $\vec{R}$. We introduce auxiliary fields $\vec{\lambda}(z,\tau)$ and write the partition function in the path-integral formalism as
\begin{equation}
\begin{split}
\mathcal{Z} &= \int{ \mathscr{D}\vec{\zeta}_0\mathscr{D}\vec{\lambda}\mathscr{D}\vec{\zeta} \: \ee^{-\{\mathcal{S} + \int\dd z\dd\tau \: \ii \vec{\lambda}(z,\tau)\cdot[\vec{\zeta}_0(z,\tau)-\vec{\zeta}(z,\vec{R},\tau)]\}\mycomment{/\hbar}} } \\
&= \int \mathscr{D}\vec{\zeta}_0\mathscr{D}\vec{\lambda}\mathscr{D}\vec{\zeta} \: \ee^{ -\sum_p\{\vec{\zeta}_p^\dag \frac{M_p}{2A_\perp\pi\beta V} \vec{\zeta}_p 
+ \ii\vec{\lambda}_{\vec{q}}^\dag \cdot [ \frac{-\ee^{-\ii\vec{Q}\cdot\vec{R}}}{\beta V} \vec{\zeta}_{p} + \frac{A_\perp}{\beta V} \vec{\zeta}_{0\vec{q}} ] \}\mycomment{/\hbar} }
\end{split}
\end{equation}
wherein we introduce a shorthand notation $\vec{q}=(q,\omega)$ for the 1D energy-momentum space. We then integrate out $\vec{\zeta}$ to obtain (omitting the determinant prefactor)
\begin{equation}
\begin{split}
\mathcal{Z} 
&= \int \mathscr{D}\vec{\zeta}_0\mathscr{D}\vec{\lambda} \: \ee^{ -\sum_{\vec{q}} [ - \frac{\pi A_\perp}{2\beta V}\vec{\lambda}_{\vec{q}}^\dag \sum_{\vec{Q}}M^{-1}_p \vec{\lambda}_{\vec{q}} 
+ \frac{\ii}{\beta\Omega}\vec{\lambda}_{\vec{q}}^\dag \cdot \vec{\zeta}_{0\vec{q}} ] \mycomment{/\hbar} }.
\end{split}
\end{equation}
Finally, by integrating out $\vec{\lambda}$, we get 
\begin{equation}
\begin{split}
\mathcal{Z} 
&= \int \mathscr{D}\vec{\zeta}_0 \: \ee^{ -\sum_{\vec{q}} \frac{1}{2\pi\beta\Omega}\frac{\Omega_\perp}{A_\perp} \vec{\zeta}_{0\vec{q}}^\dag \left( \sum_{\vec{Q}}M^{-1}_p \right)^{-1} \vec{\zeta}_{0\vec{q}}  \mycomment{/\hbar} }.
\end{split}
\end{equation}
We thus arrives at the 1D effective action
\begin{equation}\label{eq:1Daction}
\mathcal{S}_\mathrm{1D} = \frac{1}{2\pi\beta\Omega} \sum_{\vec{q}}\vec{\zeta}_{\vec{q}}^\dag \mathcal{M}_{\vec{q}} \vec{\zeta}_{\vec{q}}
\end{equation}
where $\mathcal{M}_{\vec{q}} = \frac{\Omega_\perp}{A_\perp} \left( \sum_{\vec{Q}}M^{-1}_p \right)^{-1}$, $M_p$ is given in Eq.~\eqref{eq:ChargeChiAction} and we neglect the subscript $0$ of the fields. Fortunately, this $\vec{Q}$-summation can be done analytically in the continuum limit as an integration. Using polar coordinates, the radial part of this 2D integral should be cut off at a certain $Q^*$ of the size of the 2D first Brillouin zone. In fact, it can be fixed by requiring that the effective model returns to the original model at the noninteractiong limit since the many 1D wires become completely decoupled, which simply gives ${Q^*}^2=\frac{4\pi}{A_\perp}$. The result is a block-diagonal $\mathcal{M}=\mathrm{diag}(\mathcal{M}_\theta,\mathcal{M}_\phi)$ with (omitting the subscript $\vec{q}$)
\begin{equation}\label{eq:1DactionMatr}
\begin{gathered}
\mathcal{M}_\theta = \frac{q^2}{4}
\begin{bmatrix}
v^+ & v^--\frac{2\zz}{q} \\
v^--\frac{2\zz}{q} & v^+ 
\end{bmatrix} \\
\mathcal{M}_\phi = \mycomment{\frac{\pi q^2}{Q^2A_\perp}}\frac{q^2}{4}
\begin{bmatrix}
 \frac{v^+ (q v_1-\zz) (q v_{-1}+\zz) - 2 (q v^- -2 \zz)^2 v_{g'}}{(q v_1-\zz) (q v_{-1}+\zz)-2 q^2 v^+ v_{g'}} & v^--\frac{2\zz}{q} \\
v^--\frac{2\zz}{q} &  v^+
\end{bmatrix}
\end{gathered}
\end{equation}
wherein 
\begin{equation}\label{eq:vg_renorm}
\mycomment{v_{g'}=v_g \ln{\left[1+ \frac{1}{q^2}\frac{(\zz-q v_1)(q v_{-1}+\zz)}{ \frac{1}{{Q^*}^2}\left(\zz-q v_1\right) \left(q v_{-1}+\zz\right)-2 v^+ v_g}\right]}.}
v_{g'}=v_g \ln{\left[1- \frac{(\zz-q v_1)(q v_{-1}+\zz)}{2 q^2v^+ v_g}\right]}.
\mycomment{v_{g'}=v_g \ln{\left[1- \frac{(\zz-q v_1)(q v_{-1}+\zz)}{\left(\zz-q v_1\right) \left(q v_{-1}+\zz\right)-2 q^2v^+ v_g}\right]^{-1}}.}
\end{equation}
Henceforth, we denote $v_g=\frac{g}{{Q^*}^2}=\frac{2e^2}{{Q^*}^2A_\perp}=\frac{e^2}{2\pi}$ and thereby the new $v_{g'}=\frac{g'}{{Q^*}^2}$ is understood as that the original $g$ gets renormalized to $g'$ by the logarithmic factor. 


\subsection{Localization length}\label{App:Localization}

Now we have obtained an effective TLL model Eq.~\eqref{eq:1Daction}, upon which we would consider the impurity effect. To begin with, we expand the fields in the impurity Hamiltonian around their slowly varying classical parts $\theta_\rho\rightarrow\theta_\rho^\mathrm{cl}+\theta_\rho \,, \phi_\rho\rightarrow\phi_\rho^\mathrm{cl}+\phi_\rho$. Because of the unbounded fluctuation of the fields, we have to use the normal ordering formula $\cos\varphi=:\!\cos\varphi\!:\braket{\cos\varphi}$ for a generic field $\varphi$. For a quadratic theory, the cosine product in the impurity Hamiltonian is approximated as 
\begin{equation}\label{eq:H_imp_SCHA}
\begin{split}
 \gamma [1-\frac{1}{2}(\phi_\rho^2-\braket{\phi_\rho^2} + \theta_\rho^2-\braket{\theta_\rho^2})] \cos{\phi_\rho^\mathrm{cl}}\cos{(\theta_\rho^\mathrm{cl}+\Delta kz)},
\end{split}
\end{equation}
where $\gamma$ is defined as 
\begin{equation}\label{eq:gamma}
\gamma=\ee^{-\frac{1}{2}(\braket{\phi_\rho^2} + \braket{\theta_\rho^2})}.
\end{equation}
Because of the homogeneity of the 1D spacetime, $\gamma$ is a constant as we will see below. Obviously, it introduces two mass terms to the action matrices in Eq.~\eqref{eq:1DactionMatr} in a self-consistent manner. For the compromised pinning, as implied from the impurity Hamiltonian, while $\phi_\rho^\mathrm{cl}$ should always maximize $|\cos\phi_\rho^\mathrm{cl}|$, $\theta_\rho^\mathrm{cl}$ directly affected by impurities will give rise to a coefficient of the energy gain, $-\sqrt{\frac{n_\mathrm{imp}}{L}}$, where $n_\mathrm{imp}$ is the impurity density. This is to say that the impurity Hamiltonian, using Eq.~\eqref{eq:H_imp_SCHA}, will be replaced by
\begin{equation}\label{eq:H_imp_QM}
H_\mathrm{imp} = -\sqrt{\frac{n_\mathrm{imp}}{L}} \mathcal{V}_0\gamma \int \dd z [1-\frac{1}{2}(\phi_\rho^2-\braket{\phi_\rho^2} + \theta_\rho^2-\braket{\theta_\rho^2})].
\end{equation}
 As a whole, these considerations lead to an impurity action (the two mass terms) added to Eq.~\eqref{eq:1Daction} or Eq.~\eqref{eq:1DactionMatr}
\begin{equation}\label{eq:1Daction_full}
\mathcal{S}_\mathrm{1D}' = \mathcal{S}_\mathrm{1D} + \frac{1}{2\pi\beta\Omega} \lambda \sum_{\vec{q}}(\phi_{\rho,-\vec{q}}\phi_{\rho,\vec{q}} + \theta_{\rho,-\vec{q}}\theta_{\rho,\vec{q}})
\end{equation}
wherein $\lambda = \sqrt{\frac{n_\mathrm{imp}}{L}} \mathcal{V}_0 \gamma\pi$.
In Sec.~\ref{App:solveSCeq}, the self-consistency condition Eq.~\eqref{eq:gamma} is solved to give 
\begin{equation}\label{eq:selfconEqSolution}
\gamma = \mycomment{(4\pi)^{\frac{\eta_\theta}{4-\eta}}(Q^2A_\perp)^{\frac{\eta_\phi}{4-\eta}} (4v^+\Lambda^2)^{\frac{\eta}{\eta-4}} (\mathcal{V}_0\sqrt{\frac{n_\mathrm{imp}}{L}})^{\frac{-\eta}{\eta-4}} }  (4v^+\Lambda^2)^{\frac{\eta}{\eta-4}} (4\pi \mathcal{V}_0\sqrt{\frac{n_\mathrm{imp}}{L}})^{\frac{-\eta}{\eta-4}}
\end{equation}
in which $\Lambda=\alpha^{-1}$ is the momentum cutoff, $\eta=\eta_\theta+\eta_\phi$ with $\eta_\theta = 1$, $\eta_\phi=\mycomment{\frac{\frac{Q^2A_\perp}{\pi}}{8\pi}2} \frac{1}{\pi}\tilde{F}(v_\pm,v_g)$. These are the most important exponents discussed in the main text, for which a mathematical discussion is given in Sec.~\ref{App:tildeF}.

Now it is the stage to look at the energy of this massive system due to the presence of many impurities. First of all, the penalty in elastic energy from the distortion of $\theta^\mathrm{cl}_\rho$ is estimated as\cite{SCHA1SM} $E_\mathrm{ela} = \int \dd z A_{\theta_\rho} (\nabla\theta^\mathrm{cl}_\rho)^2 = A_{\theta_\rho} \frac{\pi^2}{3L^2}\Omega$ where $A_{\theta_\rho}=\frac{1}{2\pi}\frac{v^+}{4}$. Secondly, from Eq.~\eqref{eq:1Daction_full}, we need to estimate the difference in ground state energy, $\Delta E$, between the impurity system and the original one. This usually turns out to be straightforward if one maps the bosonic action to many harmonic oscillators. However, this becomes intractable for our complex effective model Eq.~\eqref{eq:1DactionMatr} where canonical commutation relations get distorted. Instead, in Sec.~\ref{App:energyGain}, we directly calculate the free energy from the path integral of Eq.~\eqref{eq:1Daction_full}, which equals the ground state energy at zero temperature. Combining $\Delta E$ with other constant parts in Eq.~\eqref{eq:H_imp_QM}, we arrive at an energy gain $\Delta\mathcal{E}=-\sqrt{\frac{n_\mathrm{imp}}{L}} \mathcal{V}_0 \gamma\Omega
(1-\frac{\eta}{4})$. Lastly, the total energy excess density due to the impurities reads $\varepsilon=(E_\mathrm{ela}+\Delta\mathcal{E})/\Omega$, whose variation with respect to the localization length, $\frac{\partial\varepsilon}{\partial L}=0$, gives 
\begin{equation}\label{eq:L_loc}
L\propto {\mathcal{D}_\mathrm{imp}}^{\frac{1}{\eta-3}}
\end{equation}

\subsubsection{Solving the self-consistency equation}\label{App:solveSCeq}
Two positive constants $q_\theta,q_\phi$ are defined through $\lambda = \sqrt{\frac{n_\mathrm{imp}}{L}} \mathcal{V}_0 \gamma\pi = \frac{1}{4}v^+q_\theta^2 = \mycomment{\frac{\pi}{Q^2A_\perp}}\frac{1}{4}v^+q_\phi^2$. For clearness, we use different subscripts although $q_\theta=q_\phi$. From this new action, on can easily obtain the needed correlation functions
\begin{equation}\label{eq:ImpCorr}
\begin{split}
&\braket{\theta_{\rho,-\vec{q}} \theta_{\rho,\vec{q}} } = \pi\beta\Omega \frac{-v^+}{(\zz-q v_1)(q v_{-1}+\zz) -\frac{1}{4}{v^+}^2q_\theta^2} \\ 
&\braket{\phi_{\rho,-\vec{q}} \phi_{\rho,\vec{q}} } = \pi\beta\Omega \mycomment{\frac{Q^2A_\perp}{\pi}\frac{1}{v^+(q_\phi^2 + 4k^2\frac{A^2B^2}{AB-C})}}  \frac{4}{v^+(q_\phi^2 + 4k^2\frac{A^2B^2}{AB-C})}
\end{split}
\end{equation}
where for the more complex $\phi$-field case, we change to the polar coordinates defined as $\vec{k} = (q,\frac{\omega}{v^+})=(k\cos\mu,k\sin\mu)$ with $\zz=\ii\omega$ and the dimensionless functions $A=\frac{v_1}{v^+}\cos\mu-\ii\sin\mu$, $B=\frac{v_{-1}}{v^+}\cos\mu+\ii\sin\mu$ and $C=2\frac{v_{g'}}{v^+}\cos^2\mu$. For $v_{g'}$, which is originally expressed as $v_{g'} = v_g \ln[ 1 + \frac{AB}{\cos^2\mu (\frac{2v_g}{v^+} + \frac{k^2}{{Q^*}^2} AB)} ]$\mycomment{$v_{g'} = v_g \ln[ 1 - \frac{AB}{\frac{2v_g}{v^+}\cos^2\mu + ( 1 + \frac{k^2}{{Q^*}^2}\cos^2\mu) AB} ]^{-1}$}, in the low-energy regime where $\frac{k^2}{Q^{*2}}\ll 1$, we can take $v_{g'} = v_g \ln[ 1 + \frac{AB}{\frac{2v_g}{v^+} \cos^2\mu} ]$\mycomment{$v_{g'} = v_g \ln[ 1 - \frac{AB}{\frac{2v_g}{v^+} \cos^2\mu + AB} ]^{-1}$}
. Hence, $v_{g'}=v_{g'}(\mu)$ no longer depends on the variable $k$. In addition, it is convenient to define $f(\mu) = \frac{A^2B^2}{AB-C}$. The space-time correlation functions appeared in Eq.~\eqref{eq:H_imp_SCHA} and Eq.~\eqref{eq:gamma} can then calculated as follows. 
\begin{equation}\label{eq:thetamass1}
\begin{split}
\braket{\theta_\rho(z,\tau)^2} 
&= \frac{1}{(\beta\Omega)^2} \sum_{\vec{q}} \braket{\theta_{\rho,-\vec{q}} \theta_{\rho,\vec{q}} } \\
&= \frac{\pi}{\Omega} \sum_q \int_{-\infty}^\infty \frac{\dd\omega}{2\pi} \frac{-v^+}{(\zz-q v_1)(q v_{-1}+\zz) -\frac{1}{4}{v^+}^2q_\theta^2} \\
&= \frac{\pi}{\Omega} \sum_q \frac{v^+}{\sqrt{(qv_1+qv_{-1})^2+q_\theta^2{v^+}^2}} \\
&= \frac{1}{2} \int_{-\Lambda}^\Lambda \frac{1}{\sqrt{q^2+q_\theta^2}} \\
&= \ln\frac{\Lambda+\sqrt{q_\theta^2+\Lambda^2}}{\Lambda}
\end{split}
\end{equation}
in which we use the zero-temperature limit to perform the frequency summation. The momentum cutoff $\Lambda=\alpha^{-1}$, in general, is much larger than $q_\theta,q_\phi$ in the introduced masses. Therefore, if necessary, Eq.~\eqref{eq:thetamass1} can be approximated as
\begin{equation}\label{eq:thetamass2}
\braket{\theta_\rho(z,\tau)^2} = \ln\frac{2\Lambda}{q_\theta}.
\end{equation}
And for the $\phi$-field, we similarly have
\begin{equation}\label{eq:phimass1}
\begin{split}
\braket{\phi_\rho(z,\tau)^2} 
&= \frac{1}{(\beta\Omega)^2} \sum_{\vec{q}} \braket{\phi_{\rho,-\vec{q}} \phi_{\rho,\vec{q}} } \\
&= \frac{1}{2} \int \dd q \frac{\dd\omega}{2\pi} \mycomment{\frac{Q^2A_\perp}{\pi}\frac{1}{v^+(q_\phi^2 + 4k^2\frac{A^2B^2}{AB-C})}}  \frac{4}{v^+(q_\phi^2 + 4k^2\frac{A^2B^2}{AB-C})} \\
\mycomment{&= \frac{v^+}{4\pi} \int \dd^2 \vec{k} \mycomment{\frac{Q^2A_\perp}{\pi}\frac{1}{v^+(q_\phi^2 + 4k^2\frac{A^2B^2}{AB-C})}}  \frac{4}{v^+(q_\phi^2 + 4k^2\frac{A^2B^2}{AB-C})} \\}
&= \frac{v^+}{8\pi} \int_0^{2\pi} \dd\mu \int_0^\Lambda \dd k^2 \mycomment{\frac{Q^2A_\perp}{\pi}\frac{1}{v^+(q_\phi^2 + 4k^2\frac{A^2B^2}{AB-C})}}  \frac{4}{v^+(q_\phi^2 + 4k^2\frac{A^2B^2}{AB-C})} \\
&= \mycomment{\frac{1}{8\pi} \frac{Q^2A_\perp}{\pi}} \frac{1}{2\pi} \int_0^{2\pi} \dd\mu \frac{\ln(1+\frac{\Lambda^2}{q_\phi^2}4f(\mu))}{4f(\mu)} \\
&= \mycomment{\frac{1}{8\pi} \frac{Q^2A_\perp}{\pi}} \frac{1}{2\pi} F(\frac{\Lambda}{q_\phi},v_\pm,v_g).
\end{split}
\end{equation}
Again, for a very large momentum cutoff $\Lambda\gg q_\phi$, it is approximated as
\begin{equation}\label{eq:phimass2}
\braket{\phi_\rho(z,\tau)^2} = \mycomment{\frac{1}{8\pi} \frac{Q^2A_\perp}{\pi} 2} \frac{1}{\pi} \tilde{F}(v_\pm,v_g) \ln\frac{2\Lambda}{q_\phi}
\end{equation}
where $\tilde{F}(v_\pm,v_g) = \int_0^{2\pi} \dd\mu \frac{1}{4f(\mu)}$ is inspected with care in Sec.~\ref{App:tildeF}.
Feeding Eq.~\eqref{eq:thetamass2} and Eq.~\eqref{eq:phimass2} to the self-consistency equation Eq.~\eqref{eq:gamma}, we have
\begin{equation}
\gamma^{-2} = \left(\frac{2\Lambda}{q_\theta}\right)^{\eta_\theta} \left(\frac{2\Lambda}{q_\phi}\right)^{\eta_\phi}
\end{equation}
where $\eta_\theta = 1$, $\eta_\phi=\mycomment{\frac{\frac{Q^2A_\perp}{\pi}}{8\pi}2} \frac{1}{\pi}\tilde{F}$ and we also define $\eta=\eta_\theta+\eta_\phi$. Recalling the definition of $q_\theta,q_\phi$, this is further solved to give Eq.~\eqref{eq:selfconEqSolution}.

\subsubsection{Energy gain of the system}\label{App:energyGain}
Applying the path-integral formula to a generic action matrix $\mathcal{M}$
\begin{equation}
\ee^{-\beta F} = \mathcal{Z} = \int\mathscr{D}\vec{\zeta} \ee^{-\frac{1}{2\pi\beta\Omega} \sum_{\vec{q}} \vec{\zeta}_{\vec{q}}^\dag\mathcal{M}_{\vec{q}}\vec{\zeta}_{\vec{q}}},
\end{equation}
at zero temperature, energy of the system is given by
\begin{equation}\label{eq:freeenergy}
\begin{split}
E &= -\frac{1}{\beta} \ln\left[ \prod_{\vec{q}} \left(\frac{1}{(\pi\beta\Omega)^4}\mathrm{Det}\mathcal{M}_{\vec{q}}\right)^{-\frac{1}{2}} J_q \right] \\
&= \frac{1}{2} \sum_q \int\frac{\dd\omega}{2\pi} \ln [\frac{1}{(\pi\beta\Omega)^4} \mathrm{Det}\mathcal{M}_{\vec{q}}\, J_q^{-2}]
\end{split}
\end{equation}
where $J_q=q^2$ is a Jacobian since the current field variables $\vec{\zeta}$ do not directly lead to energy, dissimilar to what momentum and position do for a harmonic oscillator action. And in fact, all the factors except the determinant inside the logarithmic function will cancel out when calculating energy difference.
 
From the full action Eq.~\eqref{eq:1Daction_full}, now expressed as
\begin{equation}
\mathcal{S}_\mathrm{1D}' = \frac{1}{2\pi\beta\Omega} \sum_{\vec{q}}\vec{\zeta}_{\vec{q}}^\dag 
\begin{bmatrix}
\mathcal{M}_\theta' & 0 \\
0 & \mathcal{M}_\phi' 
\end{bmatrix}_{\vec{q}}
 \vec{\zeta}_{\vec{q}},
\end{equation}
 we can calculate the determinants 
\begin{equation}\label{eq:Det1}
\begin{split}
\mathrm{Det} \mathcal{M}_{\theta,\vec{q}}' = \frac{q^2}{16} [4(q v_1-\zz)(q v_{-1}+\zz)+{v^+}^2q_\theta^2] \\
\mathrm{Det} \mathcal{M}_{\phi,\vec{q}}' = \mycomment{(\frac{\pi}{Q^2A_\perp})^2 {v^+}^2k^2\cos^2\mu}  \frac{{v^+}^2k^2\cos^2\mu}{16} (\frac{4k^2A^2B^2}{AB-C}+q_\phi^2).
\end{split}
\end{equation}
Similarly, for the original pure system Eq.~\eqref{eq:1Daction}, we have $\mathrm{Det} \mathcal{M}_{\theta,\vec{q}}\,, \mathrm{Det} \mathcal{M}_{\phi,\vec{q}}$ by setting $q_\theta=q_\rho=0$ in Eq.~\eqref{eq:Det1}.
Then, using Eq.~\eqref{eq:freeenergy}, we have the difference in energy 
\begin{equation}
\begin{split}
\Delta E_\theta &= \frac{1}{2} \sum_q \int\frac{\dd\omega}{2\pi} (\ln [\mathrm{Det} \mathcal{M}_{\theta,\vec{q}}'] - \ln [\mathrm{Det} \mathcal{M}_{\theta,\vec{q}}] ) \\
&= \frac{1}{2} \sum_q \int\frac{\dd\omega}{2\pi} \ln \frac{4(q v_1-\zz)(q v_{-1}+\zz)+{v^+}^2q_\theta^2}{4(q v_1-\zz)(q v_{-1}+\zz)} \\
&= \frac{1}{2}\frac{\Omega}{2\pi}\int_{-\Lambda}^\Lambda\dd q  \frac{v^+}{2} (\sqrt{q^2+q_\theta^2}-|q|) \\
&= \frac{\Omega}{8\pi} v^+ [q_\theta^2 \ln\frac{\Lambda+\sqrt{q^2+\Lambda^2}}{q_\theta} + \Lambda(\sqrt{q_\theta^2+\Lambda^2}-\Lambda)].
\end{split}
\end{equation}
And for the more complex $\phi$ part, we have
\begin{equation}\label{eq:DeltaE_phi}
\begin{split}
\Delta E_\phi &= \frac{1}{2} \frac{\Omega}{2\pi}\int\dd q \int\frac{\dd\omega}{2\pi} (\ln [\mathrm{Det} \mathcal{M}_{\rho,\vec{q}}'] - \ln [\mathrm{Det} \mathcal{M}_{\rho,\vec{q}}] ) \\
&= \frac{\Omega}{(4\pi)^2} v^+ \int_0^{2\pi}\dd\mu \int_0^\Lambda\dd k^2\ln \frac{4k^2f(\mu)+q_\phi^2}{4k^2f(\mu)} \\
&= \frac{\Omega}{(4\pi)^2} v^+ \int_0^{2\pi}\dd\mu (q_\phi^2 F(\frac{\Lambda}{q_\phi},v_\pm,v_g) + Y)
\end{split}
\end{equation}
in which $Y= \int_0^{2\pi}\dd\mu \Lambda^2 \ln[1+(\frac{\Lambda^2}{q_\phi^2}4f(\mu))^{-1}]$. Similar to Eq.~\eqref{eq:thetamass2} and Eq.~\eqref{eq:phimass2}, for a large momentum cutoff $\Lambda$, we easily have $Y=q_\phi^2 \tilde{F}$.
Now, combining $\Delta E=\Delta E_\theta+\Delta E_\phi$ with other constant parts in Eq.~\eqref{eq:H_imp_QM}, i.e., using Eq.~\eqref{eq:thetamass2} and Eq.~\eqref{eq:phimass2}, we can write down the system's energy gain
\begin{equation}\label{eq:DeltaAllE}
\begin{split}
\Delta\mathcal{E} &= -\sqrt{\frac{n_\mathrm{imp}}{L}} \mathcal{V}_0\gamma \Omega [1+\frac{1}{2}(\braket{\phi_\rho^2} + \braket{\theta_\rho^2})] + \Delta E \\
&= -\sqrt{\frac{n_\mathrm{imp}}{L}} \mathcal{V}_0\gamma \Omega +  \frac{\Omega v^+}{8\pi} \Lambda(\sqrt{q_\theta^2+\Lambda^2}-\Lambda) + \frac{\Omega v^+}{(4\pi)^2} Y \\
& = -\sqrt{\frac{n_\mathrm{imp}}{L}} \mathcal{V}_0\gamma \Omega +  \frac{\Omega v^+}{8\pi} \frac{q_\theta^2}{2} + \frac{\Omega v^+}{(4\pi)^2} q_\phi^2 \tilde{F} \\
&= -\sqrt{\frac{n_\mathrm{imp}}{L}} \mathcal{V}_0 \gamma \Omega(1-\frac{\eta}{4})
\end{split}
\end{equation}
where the approximation of large $\Lambda$ is only used in the third line.

\subsubsection{$\tilde{F}$ function}\label{App:tildeF}
Let us briefly summarize the properties of the function $\tilde{F}(v_\pm,v_g) = \int_0^{2\pi} \dd\mu \frac{1}{4f(\mu)}$ introduced in Eq.~\eqref{eq:phimass2}. Recalling the definitions after Eq.~\eqref{eq:ImpCorr} in Sec.~\ref{App:solveSCeq}, $f(\mu)=\frac{A^2B^2}{AB-C}$ and $A=\frac{v_1}{v^+}\cos\mu-\ii\sin\mu$, $B=\frac{v_{-1}}{v^+}\cos\mu+\ii\sin\mu$, $C=2\frac{v_{g'}}{v^+}\sin^2\mu$ and $v_{g'} = -v_g \ln[ 1 + \frac{AB}{\frac{2v_g}{v^+} \cos^2\mu} ]$, we see it's a complex integral. Nonetheless, noticing $\Im{AB}\propto\sin{2\mu}$, it is ready to prove $f(\frac{m\pi}{2}-\mu)=f^*(\frac{m\pi}{2}+\mu)$ wherein $m\in\mathbb{Z}$, which immediately shows the reality of $\tilde{F}$ as one would expect for the physical exponent $\eta_\phi=\frac{1}{\pi}\tilde{F}$. Furthermore, $\tilde{F}$ as a bounded function of $v_g$ ($v_\pm$) is monotonically decreasing (increasing). Specifically, $\tilde{F}(\frac{v_g}{v^+}\rightarrow 0) = \pi$ and $\tilde{F}(\frac{v_g}{v^+}\rightarrow\infty) = 0$.

For the multi-copy situation, the definition of $v_{g'}$ is altered to Eq.~\eqref{eq:vg_renormN}. All the above considerations still apply except that the lower bound gets augmented to $\tilde{F}_N(\frac{v_g}{v^+}\rightarrow\infty) = \frac{N-2}{N}\pi$. Certainly, for any particular values of the arguments, $\tilde{F}_N(v_\pm,v_g)$ is larger than the single-copy one, $\tilde{F}(v_\pm,v_g)$.

\section{Temperature dependence of resistivity}\label{App:resistivity}

First of all, we need to derive the force operator used in the memory function method. Feeding the particle density Eq.~\eqref{eq:density} to the continuity equation $\nabla\cdot j+\frac{\partial\rho}{\partial t}$, we can express the current as $j=\frac{1}{\pi}\partial_t\phi_\rho$. Starting from the $\phi$-part of the noninteracting Hamiltonian $H_0^\phi=\int{\frac{\dd z}{8\pi} \left[v^+((\nabla\phi_\rho)^2 + (\nabla\phi_\chi)^2) + 2v^-(\nabla\phi_\rho)(\nabla\phi_\chi)\right]}$ obtained from Eq.~\eqref{eq:ChargeChiAction} or Eq.~\eqref{eq:1Daction} by setting $v_g=0$,
one can apply Heisenberg equation to get $j=\frac{\ii}{\pi}[H_0^\phi,\phi_\rho]$. Recalling the commutation relation Eq.~\eqref{eq:ChargeChiCommutation}, we get the current operator $j=-\frac{1}{2\pi} [v^+\nabla\phi_\chi + v^-\nabla\phi_\rho]$. Thus, using the impurity Hamiltonian, the force operator is given by
\begin{equation}
F=[j,H_\mathrm{imp}] = \ii v^+ \mathcal{V}(z)\sin\phi_\rho(z) \cos(\theta_\rho(z)+\Delta kz).
\end{equation}

To calculate the memory function $\mathsf{M}$, we need the imaginary-time force-force correlation function 
\begin{equation}
\begin{split}
\mathcal{G}(\tau) &= -\braket{\mathrm{T}_\tau F(z,\tau)F(z,0)} \\
&= {v^+}^2 \braket{\mathcal{V}\mathcal{V}} \braket{\mathrm{T}_\tau\sin{\phi_\rho}\sin{\phi_\rho}} \\
&\times\braket{\mathrm{T}_\tau\cos{(\theta_\rho+\Delta kz)}\cos{(\theta_\rho+\Delta kz)}}
\end{split}
\end{equation}
wherein we suppress the arguments for simplicity. Firstly, the Gaussian disorder correlator $\braket{\mathcal{V}(z)\mathcal{V}(z)} = \frac{1}{2}(\frac{2}{\pi\alpha})^3 \mathcal{D}_\mathrm{imp}$. For the other parts, applying the Debye-Waller formula, we have
\begin{equation}\label{eq:twoCorrForForce}
\begin{split}
&\braket{\sin{\phi_\rho}(z,\tau)\sin{\phi_\rho}(z,0)} \\
=& \frac{1}{2} \ee^{-\frac{1}{2}\braket{\left(\phi_\rho(z,\tau)-\phi_\rho(z,0)\right)^2}}  = \frac{1}{2} (v^+\Lambda\tau)^{-\eta_\phi} \\
&\braket{\cos{(\theta_\rho(z,\tau)+\Delta kz)}\cos{(\theta_\rho(z,0)+\Delta kz)}} \\
=& \frac{1}{2} \ee^{-\frac{1}{2}\braket{\left(\theta_\rho(z,\tau)-\theta_\rho(z,0)\right)^2}} = \frac{1}{2} (v^+\Lambda\tau)^{-\eta_\theta}
\end{split}
\end{equation}
wherein we suppress the time-ordering operator for simplicity. The complete force-force correlation function and the concomitant analytic continuation are formidable to obtain, especially for the complicated $\phi_\rho$ part. Nevertheless, as shown below, if one restricts the goal to only extracting the power law dependence on temperature, one can calculate the correlation functions on the exponents approximately to directly find the $\tau$ dependence as shown in Eq.~\eqref{eq:twoCorrForForce}. 

By setting $q_\theta=q_\phi=0$ in Eq.~\eqref{eq:ImpCorr}, it is straightforward to obtain the needed correlation functions in momentum space for the action Eq.~\eqref{eq:1Daction}. Then we can calculate in the following way
\begin{equation}
\begin{split}
&\braket{(\theta_\rho(z,\tau)-\theta_\rho(z,0))^2} \\
=& \frac{1}{(\beta V)^2} \sum_{\vec{q}} \braket{\theta_{\rho,-\vec{q}}\theta_{\rho,\vec{q}}} (2-2\cos\omega\tau) \\
=& \frac{\pi}{(2\pi)^2} \int \dd q\dd\omega \frac{v^+}{{v^+}^2k^2AB} (2-\cos\omega\tau) \\
=& \frac{v^+}{4\pi} \int \dd^2\vec{k} \frac{1}{v^+k^2AB} (2-2\cos{(kv^+\tau\sin\mu)})  \\
=& \frac{1}{2\pi} \int_0^{2\pi}\dd\mu\int_0^\Lambda\dd k \frac{1-\cos{(kv^+\tau\sin\mu)}}{kAB} \\
=& \frac{1}{2\pi} \int_0^{2\pi}\dd\mu \frac{1}{AB} \mathrm{Cin}(\Lambda v^+\tau\sin\mu) \\
=& \frac{1}{2\pi} \int_0^{2\pi}\dd\mu \frac{1}{AB} \ln(\Lambda v^+\tau\sin\mu) \\
=& \frac{2}{\pi} \tilde{F}(v_\pm,v_g=0) \ln(\Lambda v^+\tau) 
\end{split}
\end{equation}
and
\begin{equation}
\begin{split}
&\braket{(\phi_\rho(z,\tau)-\phi_\rho(z,0))^2} \\
=& \frac{1}{(\beta V)^2} \sum_{\vec{q}} \braket{\phi_{\rho,-\vec{q}}\phi_{\rho,\vec{q}}} (2-2\cos\omega\tau) \\
=& \frac{v^+}{4\pi} \int \dd^2\vec{k} \frac{1}{v^+k^2f(\mu)} (2-2\cos{(kv^+\tau\sin\mu)})  \\
=& \frac{1}{2\pi} \int_0^{2\pi}\dd\mu\int_0^\Lambda\dd k \frac{1-\cos{(kv^+\tau\sin\mu)}}{kf(\mu)} \\
=& \frac{1}{2\pi} \int_0^{2\pi}\dd\mu \frac{1}{f(\mu)} \mathrm{Cin}(\Lambda v^+\tau\sin\mu) \\
=& \frac{1}{2\pi} \int_0^{2\pi}\dd\mu \frac{1}{f(\mu)} \ln(\Lambda v^+\tau\sin\mu) \\
=& \frac{2}{\pi} \tilde{F}(v_\pm,v_g) \ln(\Lambda v^+\tau).
\end{split}
\end{equation}
For both of the above cases, we perform the $\vec{q}$-summation as a 2D integral in a similar manner to Eq.~\eqref{eq:phimass1}. We introduce the special cosine integral\cite{AbramowitzStegunSM} $\mathrm{Cin}(x)=\int_0^x\frac{1-\cos t}{t}\dd t$ whose asymptotic form is $\ln x$. In the last line but one we use this asymptotic form for the large cutoff $\Lambda$. We also drop the $\sin\mu$ inside the logarithm to get the last line since it doesn't contribute to the exponent of $\tau$ that relates to the temperature.

\section{Calculation for the multi-copy case}\label{App:multi-copy}

For each copy, we will have a corresponding set of $\vec{\zeta}$ fields as defined by Eq.~\eqref{eq:ChargeChiTransform1}. For the whole $\frac{N}{2}$-copy IB system ($N$ is even), we can think about the $4\frac{N}{2}\times4\frac{N}{2}$ action matrix equally divided as $\frac{N}{2}\times \frac{N}{2}$ blocks indexed by two copy indices. It comprises $\frac{N}{2}$ parts of single-copy action Eq.~\eqref{eq:ChargeChiAction} along the block-diagonal. In addition, due to the inter-copy Coulomb interaction, we get off-diagonal coupling $2q^2v_g$ between each pairs of $\phi_{\rho,\nu}\,,\phi_{\rho,\nu'}$ where $\nu\,,\nu'$ are copy indices. Then, by integrating out the wires in the same manner as in Sec.~\ref{App:Eff1Dmodel}, we arrive at the effective 1D model. 
Take all the intra-copy impurity scatterings into account amounts to self-consistently introducing mass terms to each diagonal block. 

To relieve the burden of notation, we will, when necessary, denote $\varepsilon = \left(q v_1-z\right) \left(q v_{-1}+z\right)=k^2 {v^+}^2 A(\mu) B(\mu)$, $a=A(\mu)B(\mu)$ and $b=\cos^2\mu \frac{v_{g'}}{v^+}$ in this section. First of all, upon obtaining the effective 1D model, the off-diagonal interaction between different copies will renormalize the quantity $v_g$ in a different manner compared with Eq.~\eqref{eq:vg_renorm}
\begin{equation}\label{eq:vg_renormN}
\mycomment{\mycomment{v_{g'}=v_g \ln{\left[1+ \frac{1}{q^2}\frac{(z-q v_1)(q v_{-1}+z)}{ \frac{1}{{Q^*}^2}\left(z-q v_1\right) \left(q v_{-1}+z\right)-\mycomment{2 N}Nv^+ v_g}\right]} }
v_{g'}=v_g \ln{\left[1+ \frac{1}{q^2}\frac{\varepsilon}{ \frac{\varepsilon}{{Q^*}^2}+\mycomment{2 N}Nv^+ v_g}\right]}}
\mycomment{v_{g'}=v_g \ln{\left[1- \frac{\varepsilon}{\varepsilon+\mycomment{2 N}Nq^2v^+ v_g}\right]^{-1}}}
v_{g'}=v_g \ln{\left[1+ \frac{\varepsilon}{2\frac{N}{2}q^2v^+ v_g}\right]}
\end{equation}
where $N$ appears in the denominator inside the logarithm. Accordingly, the  $\tilde{F}$ function will be altered to $\tilde{F}_N$ as stated in the main text or Sec.~\ref{App:tildeF}. Note that it is not the same as the substitution $v_g\rightarrow Nv_g$ with $\tilde{F}(v_\pm,v_g)$ becoming $\tilde{F}(v_\pm,Nv_g)$. Then we need to solve the self-consistency equation as previously done in Sec.~\ref{App:solveSCeq}. Because the various copies are on the same footing, the mass terms introduced can be taken to be the same in the first place. Therefore, it is unnecessary to keep track of the copy index unless otherwise stated. The $\braket{\theta_\rho\theta_\rho}$ correlation functions take the same form as Eq.~\eqref{eq:ImpCorr} and Eq.~\eqref{eq:thetamass1} while the $\braket{\phi_\rho\phi_\rho}$ correlation functions turn out to be rather different from Eq.~\eqref{eq:ImpCorr} and Eq.~\eqref{eq:phimass1}
\begin{equation}
\begin{split}
&\braket{\phi_{\rho,-\vec{q}} \phi_{\rho,\vec{q}} } \\
=&\pi\beta\Omega \frac{4v^+[\varepsilon(4 \varepsilon +{v^+}^2 q_{\phi }^2)-q^2 v^+ v_{g'} (8 \varepsilon +\mycomment{2 N}N {v^+}^2 q_{\phi}^2)]}{(4\varepsilon +{v^+}^2 q_{\phi}^2)[\varepsilon(4 \varepsilon +{v^+}^2 q_{\phi }^2)-\mycomment{2 N}N q^2 {v^+}^3 q_{\phi }^2 v_{g'}]} \\
=& \pi\beta\Omega \frac{4\left( 1-\frac{8k^2 b} {q_{\phi }^2 (a-\mycomment{2 N}N b)+4 k^2a} \right)}{v^+(4 k^2a+q_{\phi }^2)}.
\end{split}
\end{equation}
Then we can calculate
\begin{equation}\label{eq:phimass1N}
\begin{split}
&\braket{\phi_\rho(z,\tau)^2} \\
&= \frac{1}{(\beta\Omega)^2} \sum_{\vec{q}} \braket{\phi_{\rho,-\vec{q}} \phi_{\rho,\vec{q}} } \\
&= \frac{v^+}{8\pi} \int_0^{2\pi} \dd\mu \int_0^\Lambda \dd k^2 \frac{4\left( 1-\frac{8k^2 b} {q_{\phi }^2 (a-\mycomment{2 N}N b)+4 k^2a} \right)}{v^+(4 k^2a+q_{\phi }^2)}  \\
&= \mycomment{\frac{1}{8\pi} \frac{Q^2A_\perp}{\pi}} \frac{1}{2\pi} \int_0^{2\pi} \dd\mu \frac{1}{4a^2} \left( (a+\frac{2 b}{a-\mycomment{2 N}N b-1}) \log[1+\frac{4a\Lambda^2}{q_{\phi }^2}] \right.\\
& \left. -\frac{2b (a-\mycomment{2 N}N b)}{a-\mycomment{2 N}N b-1} \log[1+\frac{4 a \Lambda ^2}{q_{\phi }^2 (a-\mycomment{2 N}N b)}] \right).
\end{split}
\end{equation}
For a very large momentum cutoff $\Lambda\gg q_\phi$, it is approximated just as
\begin{equation}\label{eq:phimass2N}
\begin{split}
\braket{\phi_\rho(z,\tau)^2} &= \frac{1}{2\pi} \int_0^{2\pi} \dd\mu  \frac{a-2b}{4a^2} \log[\frac{4\Lambda^2}{q_{\phi }^2}] \\
&= \frac{1}{\pi} \tilde{F}_N(v_\pm,v_g) \ln\frac{2\Lambda}{q_\phi}.
\end{split}
\end{equation}

Now we turn to estimate the total energy excess of the system. Similar to what have been done in Sec.~\ref{App:energyGain}, we need the determinants of the action matrices of the whole system, $\mathcal{M}_{\vec{q}}'$ and $\mathcal{M}_{\vec{q}}$ for the massive one and the original one, respectively.
\begin{equation}
\begin{split}
&\mathrm{Det}\mathcal{M}_{\vec{q}}' = \left(\frac{q}{4}\right)^{\mycomment{2 N}N} (4 \varepsilon +{v^+}^2 q_{\theta }^2)^{\mycomment{N}\frac{N}{2}} \\ 
&\times \left(\frac{q}{4}\right)^{\mycomment{2 N}N} \left(4 \varepsilon +{v^+}^2 q_{\phi }^2\right)^{\mycomment{N}\frac{N}{2}-1} \left(\frac{4 \varepsilon ^2}{\varepsilon -\mycomment{2 N}N q^2 v^+ v_{g'}}+{v^+}^2 q_{\phi }^2\right) \\
&\mathrm{Det}\mathcal{M}_{\vec{q}} = \frac{\left(\frac{q}{2}\right)^{\mycomment{4 N}2 N} \varepsilon^{\mycomment{2 N}N+1}}{\varepsilon -\mycomment{2 N}N q^2 v^+ v_{g'}}.
\end{split}
\end{equation}
The difference in ground state energy is
\begin{equation}
\begin{split}
&\Delta E = \sum_{\vec{q}} \ln\frac{\mathrm{Det}\mathcal{M}_{\vec{q}}'}{\mathrm{Det}\mathcal{M}_{\vec{q}}} 
= \frac{v^+ \Omega}{(4\pi )^2} \left[ \mycomment{N}\frac{N}{2}(q_{\theta}^2F^0+Y^0) \right.\\
& \left. + (\mycomment{N}\frac{N}{2}-1)(q_{\phi}^2F^0+Y^0)+(q_{\phi }^2 F(Nv_g)+Y(Nv_g)) \right].
\end{split}
\end{equation}
And the other part from the constant terms reads
\begin{equation}
\begin{split}
E' &= -\mycomment{N}\frac{N}{2}\sqrt{\frac{n_\mathrm{imp}}{L}} \mathcal{V}_0\gamma \Omega [1+\frac{1}{2}(\braket{\phi_\rho^2} + \braket{\theta_\rho^2})] \\
&= - \mycomment{N}\frac{N}{2} \left( \sqrt{\frac{n_\mathrm{imp}}{L}} \mathcal{V}_0\gamma \Omega + \frac{v^+\Omega}{(4\pi)^2} \left( q_{\theta}^2 F^0 + q_{\phi}^2 F_N(v_g) \right) \right).
\end{split}
\end{equation}
Here, $F$ and $Y$ follow the definitions in Eq.~\eqref{eq:phimass1} and Eq.~\eqref{eq:DeltaE_phi} and we omit the first slot of arguments $v_\pm$. $F^0$ or $Y^0$ simply means setting the second argument $v_g$ to zero. And the subscript of $F_N$ means it uses the multi-copy $v_{g'}$ in Eq.~\eqref{eq:vg_renormN} as the same as $\tilde{F}_N$ introduced in Sec.~\ref{App:tildeF}. Similar to the spirit of Sec.~\ref{App:Localization}, if we approximate these functions at a very large cutoff $\Lambda$, after some lengthy manipulations, we arrive at the concise expression 
\begin{equation}
\Delta\mathcal{E} = \Delta E + E' = -\mycomment{N}\frac{N}{2}\sqrt{\frac{n_\mathrm{imp}}{L}} \mathcal{V}_0 \gamma \Omega(1-\frac{\eta_N}{4})
\end{equation}
which shares the same form as Eq.~\eqref{eq:DeltaAllE} and hence guarantees the same conclusion as Eq.~\eqref{eq:L_loc} with a new exponent $\eta_N = 1+\frac{1}{\pi}\tilde{F}_N(v_\pm,v_g)$.

\end{document}